\documentclass[a4paper,11pt]{article}
\usepackage[utf8]{inputenc}
\usepackage{lmodern}
\usepackage[T1]{fontenc}
\usepackage{amsmath,amsthm,amsfonts,amssymb,bm}
\usepackage{dsfont}			% indicator function
\usepackage{enumitem}
\usepackage{graphicx}
\usepackage{bbm}
\usepackage{booktabs}
\usepackage{xcolor}
\usepackage{authblk}
\usepackage[authoryear]{natbib}
\usepackage[right=2.5cm,left=2.5cm,top=2cm,bottom=3cm]{geometry}
\usepackage[font={small,it}]{caption}	% the options makes the font italic in floats
\usepackage[colorlinks,linkcolor=blue,citecolor=blue,urlcolor=blue]{hyperref}
\usepackage{subfig}
\usepackage{mathtools}

\usepackage{multirow}
\usepackage{makecell}
\usepackage{afterpage}

\title{ \textsc{Co-clustering of time-dependent data \\via Shape Invariant Model} }
\author[1]{Alessandro Casa}
\author[2]{Charles Bouveyron}
\author[2,3]{Elena Erosheva}
\author[4]{Giovanna Menardi}
\affil[1]{School of Mathematics \& Statistics and Vistamilk SFI Research Centre, University College Dublin}
\affil[2]{Université C\^ote d'Azur, INRIA, CNRS, Laboratoire J.A. Dieudonné, MAASAI research team}
\affil[3]{Department of Statistics, University of Washington}
\affil[4]{Deparment of Statistical Sciences, University of Padova
} 
\date{}                     %% if you don't need date to appear

\begin{document}

\maketitle

\begin{abstract}
Multivariate time-dependent data, where multiple features are observed over time for a set of individuals, are increasingly widespread in many application domains. To model these data we need to account for relations among both time instants and variables and, at the same time, for subjects heterogeneity. 
We propose a new co-clustering methodology for clustering individuals and variables simultaneously that is designed to handle both functional and longitudinal data.
% In order to sort out such complexities, we propose a new co-clustering methodology specifically tailored to the given context, allowing to cluster individuals and variables, and designed to handle both functional and longitudinal data. 
Our approach borrows some concepts from the \emph{curve registration} framework by embedding the \emph{Shape Invariant Model} in the \emph{Latent Block Model}, estimated via a suitable modification of the SEM-Gibbs algorithm. The resulting procedure allows for several user-defined specifications of the notion of cluster that could be chosen on substantive grounds and provides parsimonious summaries of complex longitudinal or functional data by partitioning data matrices into homogeneous blocks.
%provides parsimonious summaries of complex longitudinal or functional data by partitioning data matrices into homogeneous blocks and allows for different specification of the cluster notion. %In order to estimate the model, we propose an appropriate modification of the SEM-Gibbs algorithm. %The choice concerning the number of row and column clusters is carried out by means of an approximate version of the ICL criterion.
\end{abstract}

\smallskip
\noindent \textbf{Keywords:} co-clustering, curve registration, latent block model, stochastic EM

%-------------------------------------------------------------------------
\section{Introduction} \label{sec:intro}
Time dependent data, arising when measurements are taken on a set of units at different time occasions, are pervasive in a plethora of different fields. Examples include, but are not limited to, time evolution of asset prices and volatility in finance, growth of countries as measured by economic indices, heart or brain activities as monitored by medical instruments, disease evolution evaluated by suitable bio-markers in epidemiology. When analyzing such data, we need strategies modelling typical time courses by accounting for the individual correlation over time.
In fact, while nomenclature and taxonomy in this setting are not always consistent, we might highlight some relevant differences, subsequently implying different challenges in the modelling process, in time-dependent data structures. On opposite poles, we may distinguish functional from longitudinal data analysis. In the former case the quantity of interest is supposed to vary over a continuum and usually a large number of regularly sampled observations is available, allowing to treat each element of the sample as a function. On the other hand, in longitudinal studies, time observations are often shorter with sparse and irregular measurements. Readers may refer to \citet{rice2004functional} for a thorough comparison and discussion about differences and similarities between functional and longitudinal data analysis.

Early development in these areas mainly aimed to describe individual-specific curves by properly accounting for the correlation between measurements for each subject \citep[see e.g.][and references therein]{diggle2002analysis,ramsey2005functional} with the subjects themselves often considered to be independent. 
Nonetheless this is not always the case. Therefore, more recently, there has been an increased attention towards clustering methodologies aimed at describing heterogeneity among time-dependent observed trajectories;
%possible heterogeneities and relationships among the observed curves; 
see \citet{erosheva2014breaking} for a recent review of related methods used in criminology and developmental psychology. 
% Here cluster analysis is particularly useful since it accounts for heterogeneous behaviors by assuming the presence of some group-specific generative models. 
%% EE: The above sentence seems to be redundant.
From a functional standpoint, different approaches have been studied and readers may refer to the works by \citet{bouveyron2011model}, \citet{bouveyron2015discriminative} and \citet{bouveyron2020co} or to \citet{jacques2014functional} for a review. On the other hand, from a longitudinal point of view, model-based clustering approaches have been proposed by \citet{de2008model}, \citet{mcnicholas2010model}. Lastly a review from a more general time-series data perspective may be found in \citet{liao2005clustering} and \citet{frauhwirth2011model}.

\begin{figure}[t]
	\centering
	\includegraphics[width = 9cm, height = 8cm]{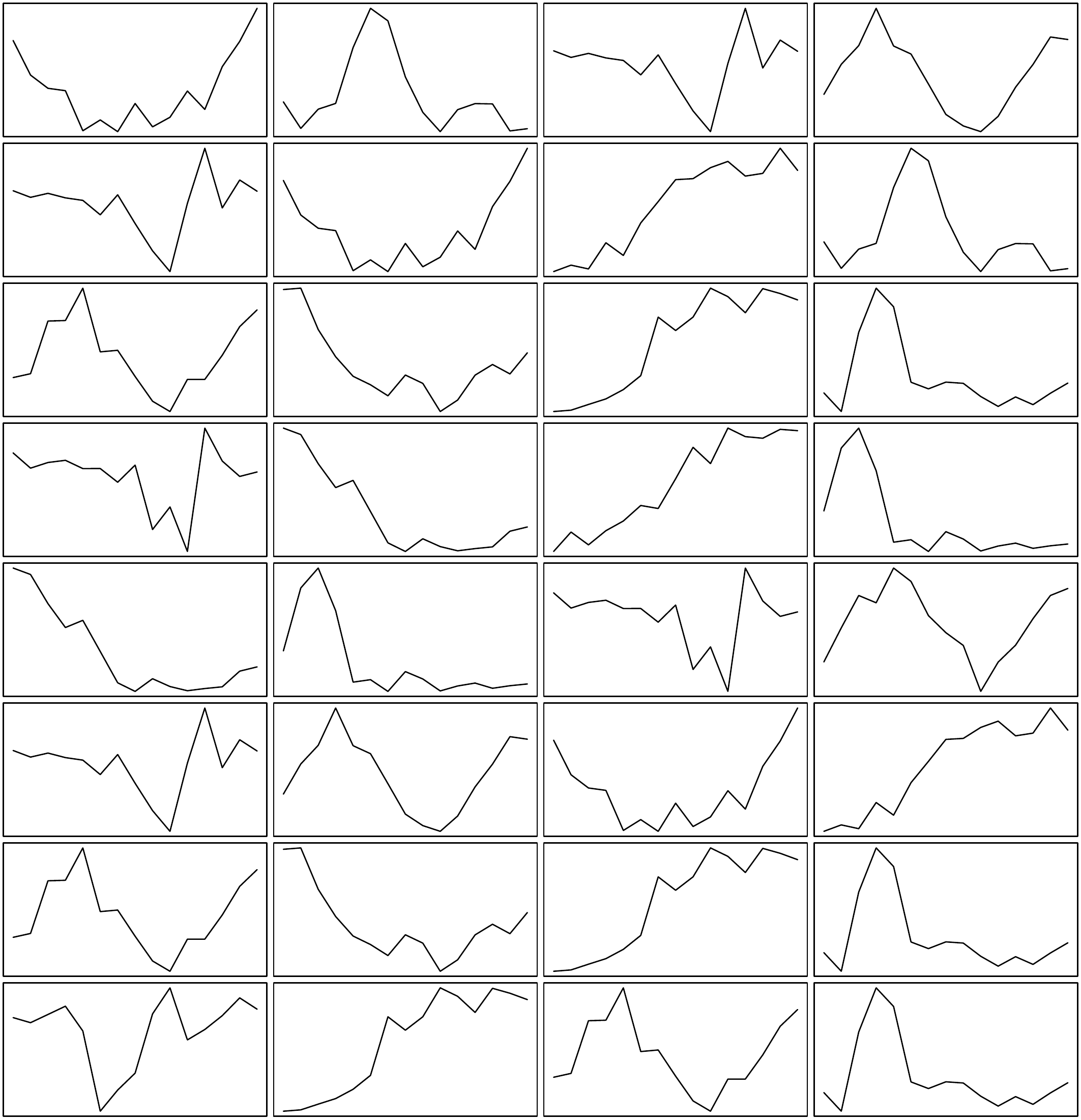}
	\caption{Example of multivariate time-dependent data: $d=4$ variables are measured for $n=8$ individuals over $T=15$ time instants, giving rise to the displayed curves.}
	\label{fig:raw_curves_multiple}
\end{figure}

The methods cited so far usually deal with situations where a single feature is measured over time for a number of subjects, where therefore the data are represented by a $n \times T$ matrix, with $n$ and $T$ being the number of subjects and of observed time occasions. Nonetheless it is increasingly common to encounter multivariate time-dependent data, where several variables are measured over time for different individuals. These data may be represented according to three-way $n \times d \times T$ matrices, with $d$ being the number of time-dependent features; a graphical illustration of such structure is displayed in Figure  \ref{fig:raw_curves_multiple}. The introduction of an additional layer entails new challenges that have to be faced by clustering tools. In fact, as noted by \citet{anderlucci2015covariance}, models have to account for three different aspects, being the correlation across different time observations, the relationships between the variables and the heterogeneity among the units, each one of them arising from a different layer of the three-way data structure.

%% EE: As I indecated before, this quote is not clear. See my comments on the earlier draft. 

To extract useful information and unveil patterns from such complex structured and high-dimensional data, standard clustering strategies would require specification and estimation of severely parameterized models. In this situation, parsimony has often been induced by somehow neglecting the correlation structure among variables. A possible clever workaround, specifically proposed in a parametric setting, is represented by the contributions of \citet{viroli2011finite,viroli2011model} where, in order to handle three-way data, mixtures of Gaussian matrix-variate distributions are exploited. 

In the present work, a different direction is taken, and a co-clustering strategy is pursued to address the mentioned issues. The term co-clustering refers to those methods aimed to find simultaneously row and column clusters of a data matrix. These techniques are particularly useful in high-dimensional settings where standard clustering methods may fall short in uncovering meaningful and interpretable row groups because of the high number of variables. By searching for homogeneous blocks in large matrices, co-clustering tools provide parsimonious summaries possibly useful both as dimensionality reduction and as exploratory steps. These techniques are particularly appropriate when relations among the observed variables are of interest.  Note that, even in the co-clustering context, the usual dualism between \emph{distance-based} and \emph{density-based} strategies can be found. We pursue the latter approach which embeds co-clustering in probabilistic framework, builds a common framework to handle different types of data and reflects the idea of a density resulting from a mixture model.  
% being partitionable in several blocks. 
Specifically, we propose a parametric model for time-dependent data and a new estimation strategy to handle the distinctive characteristics of the model. Parametric co-clustering of time-dependent data has been pursued by  \citet{slimen2018model} and \citet{bouveyron2018functional} in the functional setting, by mapping the original curves to the space spanned by the coefficients of a basis expansion. Unlike these contributions, we build on the idea that individual curves within a cluster arise as transformations of a common shape function which is modeled in such a way as to handle both functional and longitudinal data. 
%% EE: Not clear what is "modeled"? The shape function? The model? How is it "modeled" to handle both types of data?
% The considered 
Our co-clustering framework allows for easy
% us to get a notable gain not only in terms of 
interpretation and cluster description but also 
% in terms of flexibility, as it embeds different 
for specification of different notions of clusters which, depending on subject matter application, may be more desirable and interpretable by subject matter experts.   
% suitable to be assumed.

The rest of the paper is organized as follows. In Section \ref{sec:chcharles_buildingblocks}, we provide the background needed for the specification of the proposed model which is described in Section \ref{sec:chcharles_timedepLBM}, along with the estimation procedure. In Section \ref{sec:chcharles_numexample}, the model performances are illustrated both on simulated and real examples. We conclude the paper by summarizing our contributions and pointing to some future research directions.
% Some remarks conclude the paper.

%%%%%%%%%%%%%%%%%%%%%%%%%%%%%%%%%%%%%%%%%%%%%%%%%%%%%%%%%%%%%%%%%

\section{Modelling time-dependent data}\label{sec:chcharles_buildingblocks}
When dealing with the heterogeneous time dependent data landscape, outlined in the previous section, a variety of modelling approaches are sensible to be pursued. 
The route we follow in this work borrows its rationale from the \emph{curve registration} framework \citep{ramsay1998curve}, according to which observed curves often exhibit common patterns but with some variations. Methods for curve registration, also known as \emph{curve alignment} or \emph{time warping}, are based on the idea of aligning prominent features in a set of curves via either an \emph{amplitude variation}, a \emph{phase variation} or a combination of the two. The first one is concerned with vertical variations while the latter regards horizontal, hence time related, ones. As an example it is possible to think about modelling the evolution of a specific disease. Here the observable heterogeneity of the raw curves can often be disentangled in two distinct sources: on the one hand, it could depend on differences in the intensities of the disease among subjects whereas, on the other hand, there could be different ages of onset, i.e. the age at which an individual experiences the first symptoms. After properly taking into account these causes of variation, the curves result to be more homogeneously behaving, with a so called \emph{warping function}, which synchronizes the observed curves and allows for visualization and estimation of a common mean shape curve. 

Coherently with the aforementioned rationale, in this work time dependency is accounted for via a \emph{self-modelling regression} approach \citep{lawton1972self} and, more specifically, via an extension to the so called  \emph{Shape Invariant Model} \citep[SIM,][]{lindstrom1995self},  based on the idea that an individual curve arises as a simple transformation of a common shape function. \\
Let $\mathcal{X}=\{x_i({\bf t_i})\}_{1\le i \le n}$ be the set of curves, observed on $n$ individuals, with $x_i(t)$ being the level of the \emph{i-th} curve at time $t$ and $t \in {\bf t_i} = (t_1, \dots, T_{n_i})$, hence with the number of observed measurements allowed to be subject-specific. Stemming from the SIM, $x_i(t)$  is modelled as 
\begin{eqnarray}\label{eq:shapeinvariantmodel_nocluster}
x_i(t) = \alpha_{i,1} + \text{e}^{\alpha_{i,2}}m(t-\alpha_{i,3}) + \epsilon_{i}(t) \; \end{eqnarray} 
where 
\begin{itemize}
\item $m(\cdot)$ denotes a general common shape function whose specification is arbitrary. In the following we consider B-spline basis functions \citep{de1978practical}, i.e. giving $m(t)=m(t;\beta)= \mathcal{B}(t)\beta,$ where $\mathcal{B}(t)$ and $\beta$ are respectively a vector of B-spline basis evaluated at time $t$ and a vector of basis coefficients whose dimensions allow for different degrees of flexibility;
\item  $\alpha_{i}=(\alpha_{i,1},\alpha_{i,2},\alpha_{i,3}) \sim \mathcal{N}_3(\mu^\alpha,\Sigma^{\alpha})$ for $i=1,\dots,n$ is a vector of subject-specific normally distributed random effects. These random effects are responsible for the individual specific transformations of the mean shape curve $m(\cdot)$ assumed to generate the observed ones. In particular $\alpha_{i,1}$ and $\alpha_{i,3}$ govern respectively amplitude and phase variations while $\alpha_{i,2}$ describes possible scale transformations. Random effects also allow accounting for the correlation among observations on the same subject measured at different time points. Following \citet{lindstrom1995self}, the parameter $\alpha_{i,2}$ is here optimized in the log-scale to avoid identifiability issues;
\item $\epsilon_{i}(t) \sim \mathcal{N}(0,\sigma^2_{\epsilon})$ is a Gaussian distributed error term.
\end{itemize}

Due to its flexibility, the SIM has already been considered as a stepping stone to model functional as well as longitudinal time-dependent data
%different types of time-dependent data as functional and longitudinal ones 
\citep{telesca2008bayesian,telesca2012modeling}. Indeed, on the one hand, the smoothing involved in the specification of $m(\cdot;\beta)$ allows to handle function-like data. On the other hand, random effects, which borrow information across curves, make this approach fruitful even with short, irregular and sparsely sampled time series; readers may refer to \citet{erosheva2014breaking} for an illustration of this capabilities in the context of behavioral trajectories. Therefore, we find model (\ref{eq:shapeinvariantmodel_nocluster}) particularly appealing and suitable for our aims, potentially able to handle time-dependent data in a comprehensive way.  

%%%%%%%%%%%%%%%%%%%%%%%%%%%%%%%%%%%%%%%%%%%%%%%%%%%%%%%%%%%%

\section{Time-dependent Latent Block Model}\label{sec:chcharles_timedepLBM}
\subsection{Latent Block Model}\label{sec:chcharles_lbmgeneral}
In the parametric, or model-based, co-clustering framework, the \emph{Latent Block Model} \citep[LBM,][]{govaert2013co} is the most popular approach. Data are represented in a matrix form $\mathcal{X}=\{ x_{ij} \}_{1\le i \le n, 1 \le j \le d}$, where by now we should intend $x_{ij}$ as a generic random variable. To aid the definition of the model, and in accordance with the parametric approach to clustering \citep{fraley2002model,bouveyron2019model}, two latent random vectors $\mathbf{z} = \{ z_{i}\}_{1 \le i \le n}$, and $\mathbf{w}=\{ w_{j} \}_{1 \le j \le d}$, with $z_i = (z_{i1},\dots,z_{iK})$, $w_j=(w_{j1},\dots,w_{jL})$, are introduced, indicating respectively the row and column memberships, with $K$ and $L$ the number of row and column clusters. A standard binary notation is used for the latent variables, i.e. $z_{ik}=1$ if the \emph{i-th} observation belongs to the \emph{k-th} row cluster and 0 otherwise and, likewise, $w_{jl}=1$ if the \emph{j-th} variable belongs to the \emph{l-th} column cluster and 0 otherwise. The model formulation relies on a local independence assumption, i.e. the $n \times d$ random variables $\{ x_{ij} \}_{1\le i \le n, 1 \le j \le d}$ are assumed to be independent conditionally on $\mathbf{z}$ and $\mathbf{w},$ in turn supposed to be independent. The LBM can be thus written as
\begin{eqnarray}\label{eq:LBM}
p(\mathcal{X}; \Theta) = \sum_{z \in Z}\sum_{w \in W}p(\mathbf{z};\Theta)p(\mathbf{w};\Theta)p(\mathcal{X}|\mathbf{z},\mathbf{w};\Theta) \; , 
\end{eqnarray}
where
\begin{itemize}
  \item $Z$ and $W$ are the sets of all the possible partitions of rows and columns respectively in $K$ and $L$ groups; 
  \item the latent vectors $\mathbf{z}, \mathbf{w}$ are assumed to be multinomial, with $p(\mathbf{z};\Theta)=\prod_{ik}\pi_k^{z_{ik}}$ and $p(\mathbf{w};\Theta)=\prod_{jl} \rho_l^{w_{jl}}$ where $\pi_k, \rho_l > 0$ are the row and column mixture proportions, $\sum_k \pi_k = \sum_l \rho_l = 1$; 
  \item as a consequence of the local independence assumption, $p(\mathcal{X}|\mathbf{z},\mathbf{w};\Theta) = \prod_{ijkl} p(x_{ij};\theta_{kl})^{z_{ik}w_{jl}}$ where $\theta_{kl}$ is the vector of parameters specific to block $(k,l)$;  \item $\Theta = (\pi_k,\rho_l,\theta_{kl})_{1\le k \le K, 1 \le l \le L}$ is the full parameter vector of the model. 
\end{itemize}
%It is straightforward to note, from the formulation outlined in (\ref{eq:LBM}), how the introduction of an additional latent variable basically adds a supplementary mixture layer to the standard formulation of the mixture models used as a cornerstone in parametric clustering. For a more detailed tractation of the link among the LBM and mixture models the reader may refer to \citet{govaert2013co}. The authors indeed highlight how, conditionally on the partition $\mathbf{w}$, the density function of $\mathcal{X}$ is a mixture model, and the same holds conditioning on $\mathbf{z}$.

The LBM is particularly flexible in modelling different data types, as handled by a proper specification of the marginal density $p(x_{ij};\theta_{kl})$ for binary \citep{govaert2003clustering}, count \citep{govaert2010latent}, continuous \citep{lomet2012selection}, categorical \citep{keribin2015estimation}, ordinal \citep{jacques2018model,corneli2019co}, and even mixed-type data \citep{selosse2020model}.% To the best of our knowledge the only works proposing a parametric co-clustering approach to model time-dependent data are the ones of \citet{slimen2018model} and \citet{bouveyron2018functional} where the LBM is extended to a functional setting. 
%These works adopt a different, two steps, rationale with respect to the one we follow: in the first step the original curves are mapped to the space spanned by the coefficients of a basis expansion, in the second one Gaussian co-clustering is performed directly on those coefficients. As it will be clarified later on, we conversely specify a suitable generating model for the curves with a possible benefit in terms of interpretation.

\subsection{Model specification}\label{sec:chcharles_modspec}
Once the LBM structure has been properly defined, extending its rationale to handle time-dependent data in a co-clustering framework boils down to a suitable specification of $p(x_{ij};\theta_{kl})$. Note that this reveals one of the main advantage of such an highly-structured model, consisting in the chance to search for patterns in multivariate and complex data by specifying only the model for the variable $x_{ij}$. 
As introduced in Section \ref{sec:intro}, multidimensional time-dependent data may be represented according to a three-way structure where the third \emph{mode} accounts for the time evolution. The observed data assume an array configuration $\mathcal{X}= \{ x_{ij}({\bf t_i}) \}_{1\le i \le n, 1\le j \le d}$ with ${\bf t_i}=(t_1,\dots,T_{n_i})$ as outlined in Section \ref{sec:chcharles_buildingblocks}; different observational lengths can be handled by a suitable use of missing entries. Consistently with (\ref{eq:shapeinvariantmodel_nocluster}), we consider as a generative model for the curve in the $(i,j)$\emph{-th} entry, belonging to the generic block $(k,l)$, the following 
\begin{eqnarray}\label{eq:simmodel_cluster}
x_{ij}(t)|_{z_{ik}=1,w_{jl}=1} = \alpha_{ij,1}^{kl} + \text{e}^{\alpha_{ij,2}^{kl}}m(t-\alpha_{ij,3}^{kl}; \beta_{kl}) + \epsilon_{ij}(t) \; 
\end{eqnarray}
with $t \in {\bf t_i}$ a generic time instant.
A relevant difference with respect to the original SIM consists, coherently with the co-clustering setting, in the parameters being block-specific since the generative model is specified conditionally to the block membership of the cell. As a consequence:
\begin{itemize}
\item $m(t;\beta_{kl})= \mathcal{B}(t)\beta_{kl}$ where the quantities are defined as in Section \ref{sec:chcharles_buildingblocks}, with the only difference that $\beta_{kl}$ is a vector of block-specific basis coefficients, hence allowing different mean shape curves across different blocks; 
\item  $\alpha_{ij}^{kl}=(\alpha_{ij,1}^{kl},\alpha_{ij,2}^{kl},\alpha_{ij,3}^{kl}) \sim \mathcal{N}_3(\mu_{kl}^\alpha,\Sigma_{kl}^{\alpha})$ is a vector of cell-specific random effects distributed according to a block-specific Gaussian distribution;
\item $\epsilon_{ij}(t) \sim \mathcal{N}(0,\sigma^2_{\epsilon,kl})$ is the error term distributed as a block-specific Gaussian; 
\item  $\theta_{kl}=(\mu_{kl}^\alpha,\Sigma_{kl}^{\alpha},\sigma^2_{\epsilon,kl},\beta_{kl})$.
\end{itemize}
 
\begin{figure}[t]
\center
\includegraphics[width=.62\textwidth]{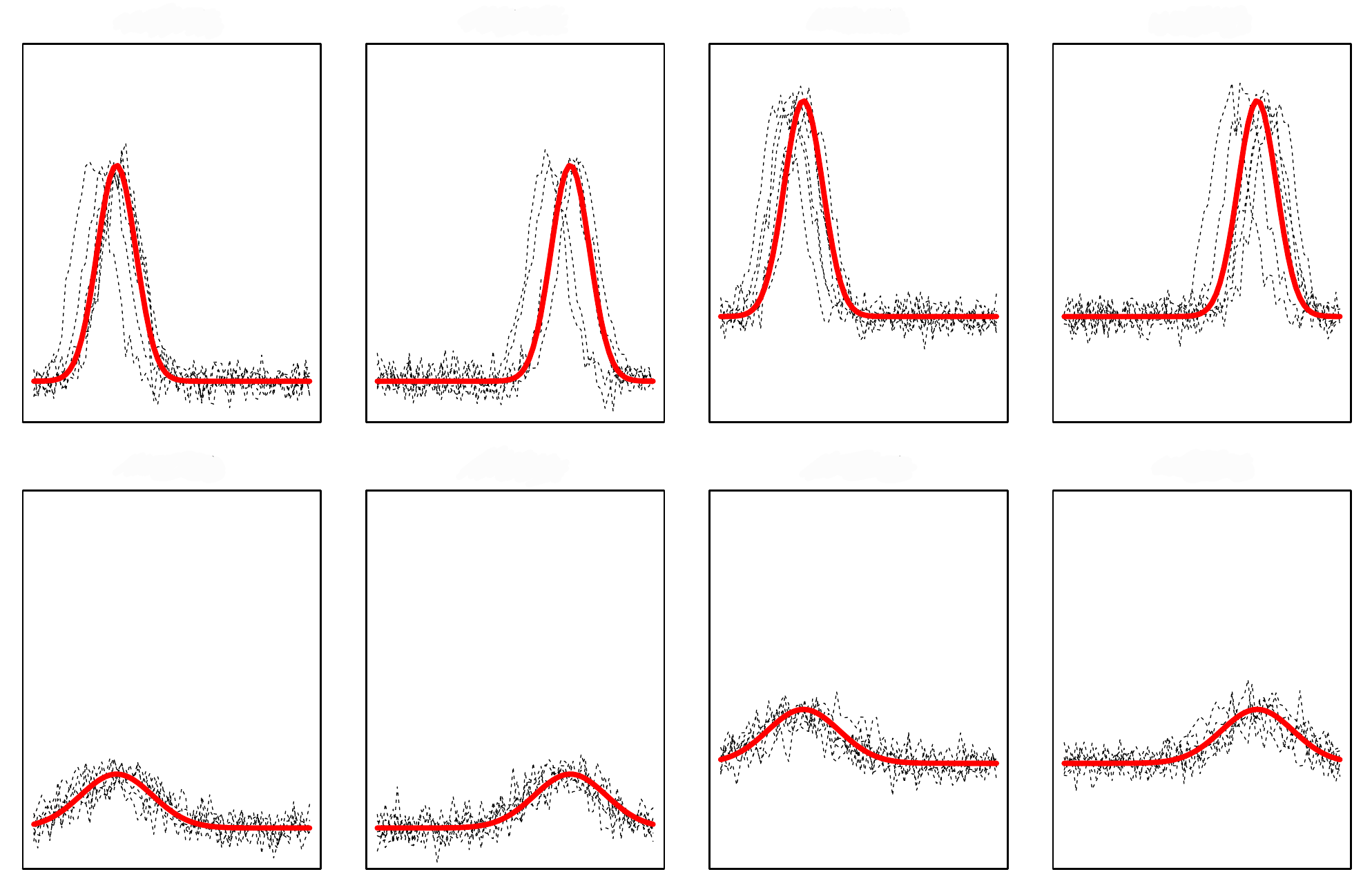}
\raisebox{2.5ex}{
\includegraphics[width=.34\textwidth]{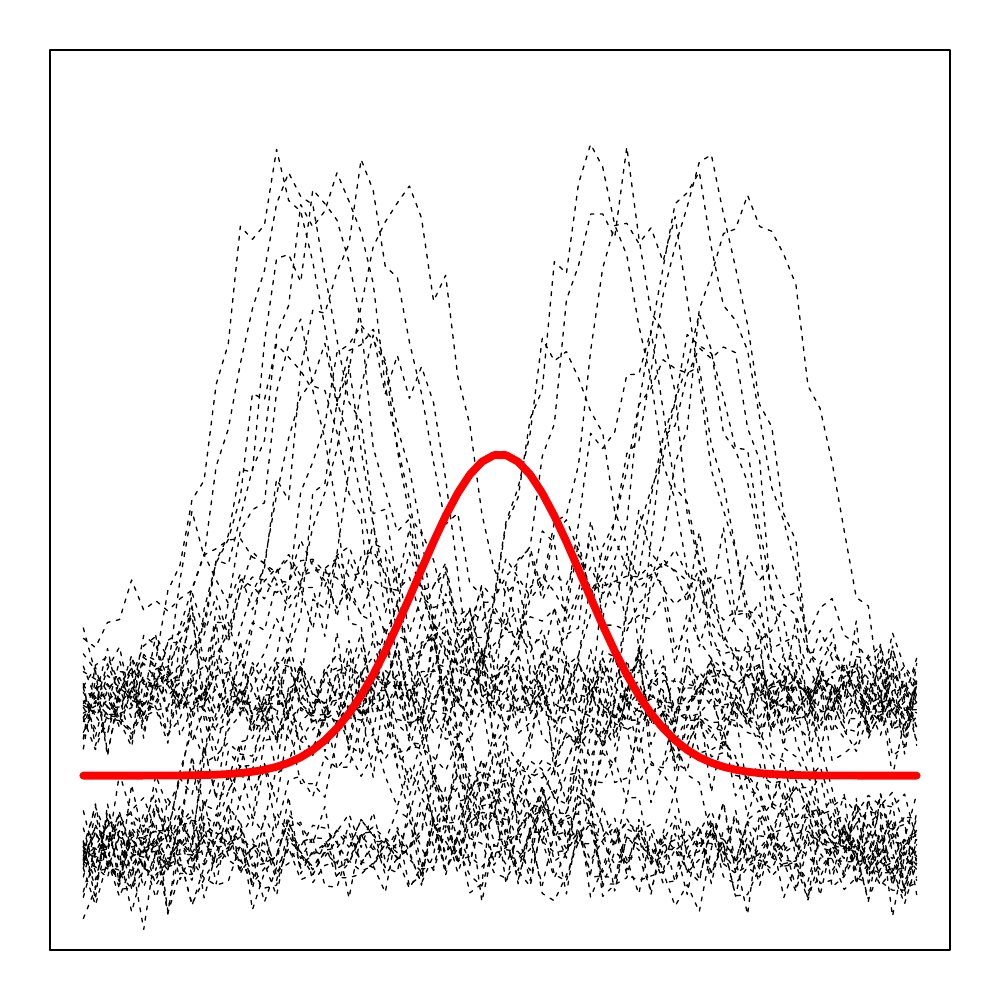}}
\caption{In the left panels, curves in dotted line arise as random fluctuations of the superimposed red curve, but they are all time, amplitude or scale transformations of the same mean-shape function on the right panel.}
\label{fig:curve_coclust}
\end{figure}

Note that the ideas borrowed from the \emph{curve registration} framework are here embedded in a clustering setting. Therefore, while \emph{curve alignment} aims to synchronize the curves to estimate a common mean shape, in our setting the SIM works as a suitable tool to model the heterogeneity inside a block and to introduce a flexible notion of cluster. The rationale behind considering the SIM in a co-clustering framework consists in looking for blocks characterized by a different mean shape function $m(\cdot;\beta_{kl})$. Curves within the same block arise as random shifts and scale transformations of $m(\cdot;\beta_{kl}),$ driven by the block-specifically distributed random effects. Let consider the small panels on the left side of Figure \ref{fig:curve_coclust}, displaying a number of curves which arise as transformations induced by non-zero values of $\alpha_{ij,1},$ $\alpha_{ij,2}, $ or $\alpha_{ij,3}$. Beyond the sample variability, the curves differ for a (phase) random shift on the $x-$axes, an amplitude shift on the $y-$ axes, and a scale factor. According to model (\ref{eq:simmodel_cluster}), all those curves belong to the same cluster, since they share the same mean shape function (Figure \ref{fig:curve_coclust}, right panel). 

\begin{figure}[t]
\center
\begin{tabular}{p{3cm}p{4cm}p{4cm}p{4cm}}
&\texttt{FTT}&\hspace{-0.35cm}\texttt{TFT}&\hspace{-0.65cm}\texttt{TTF}\\
\end{tabular}\\
\includegraphics[width=.25\textwidth]{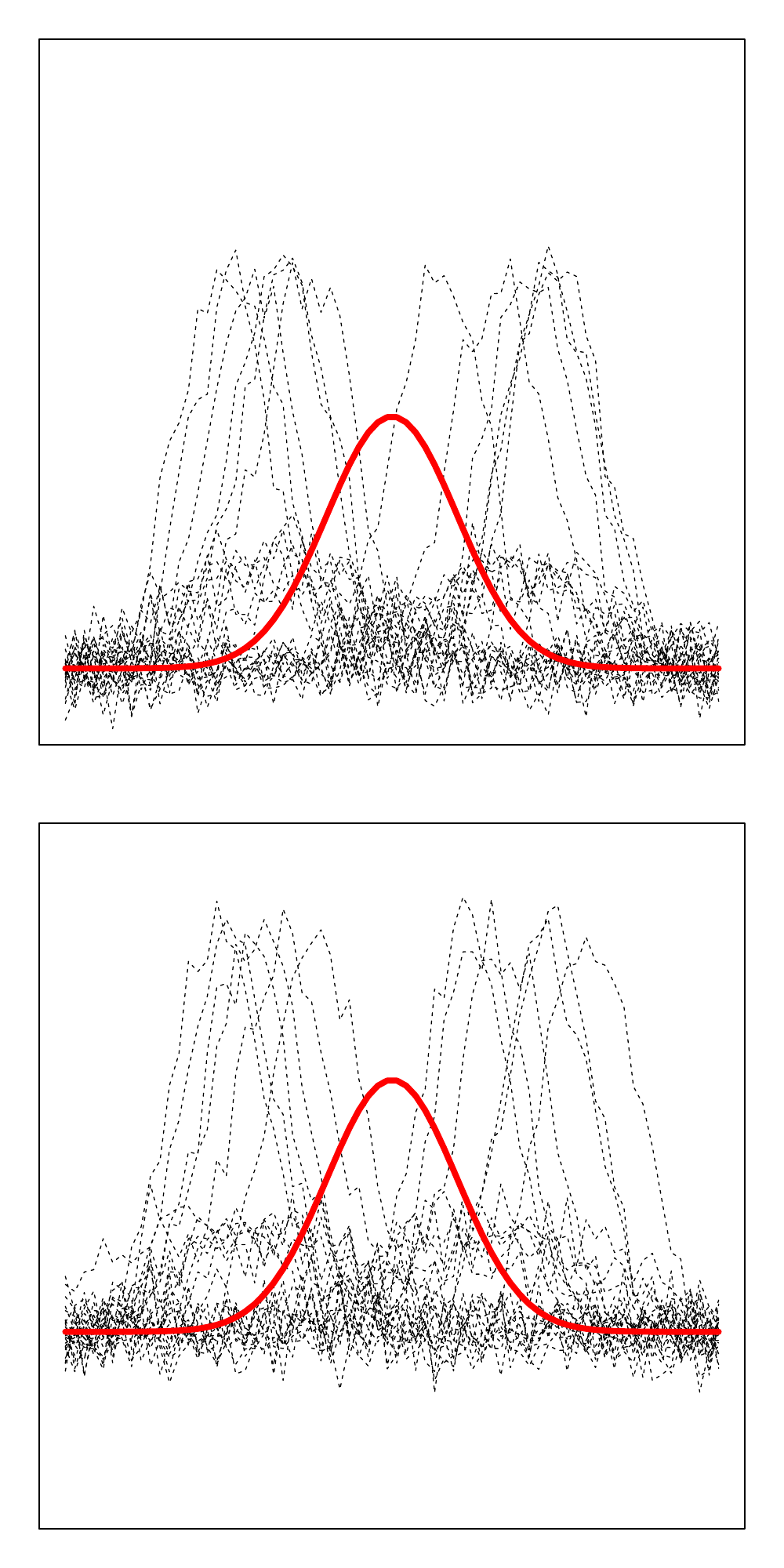}
\includegraphics[width=.25\textwidth]{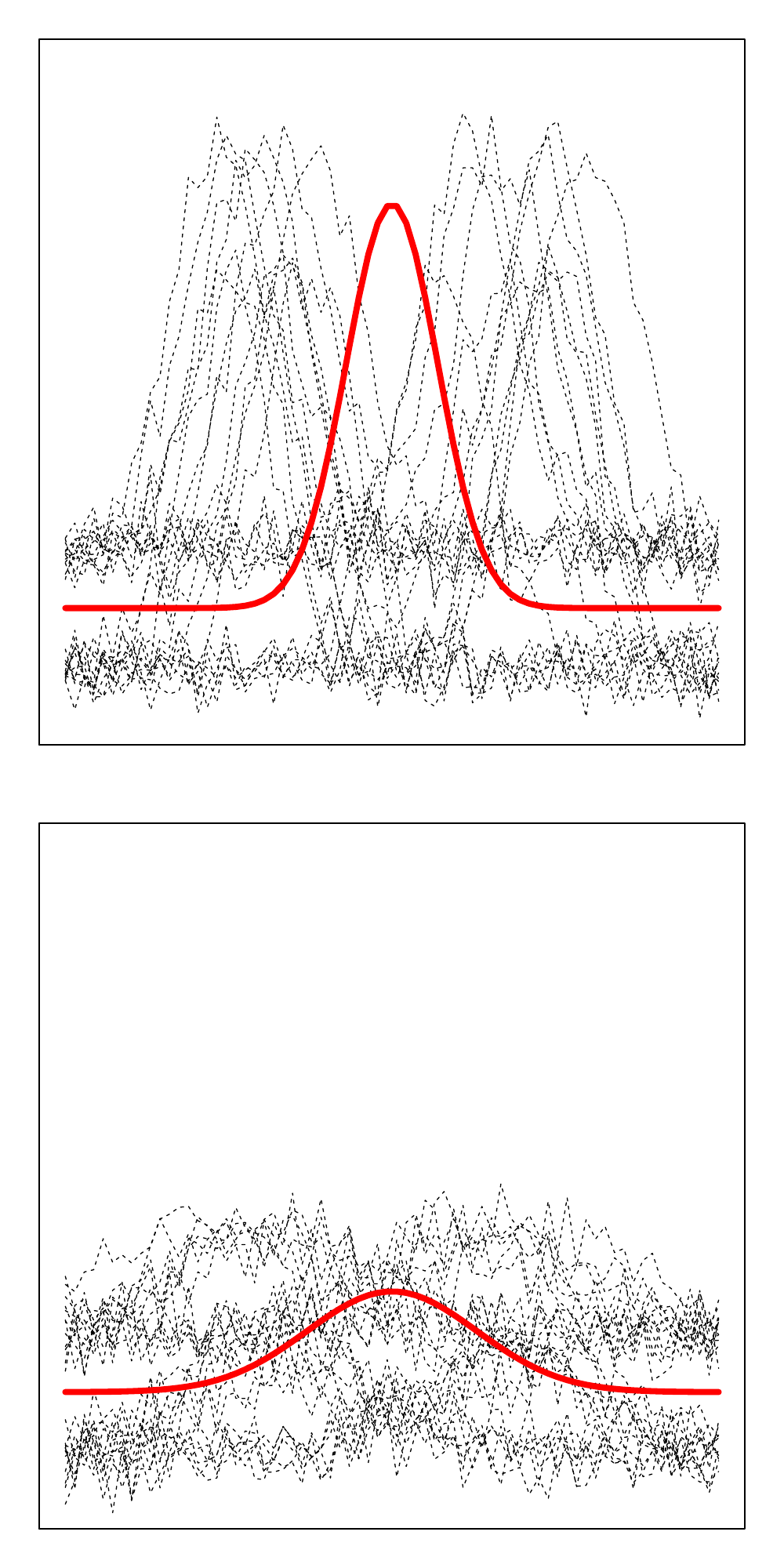}
\includegraphics[width=.25\textwidth]{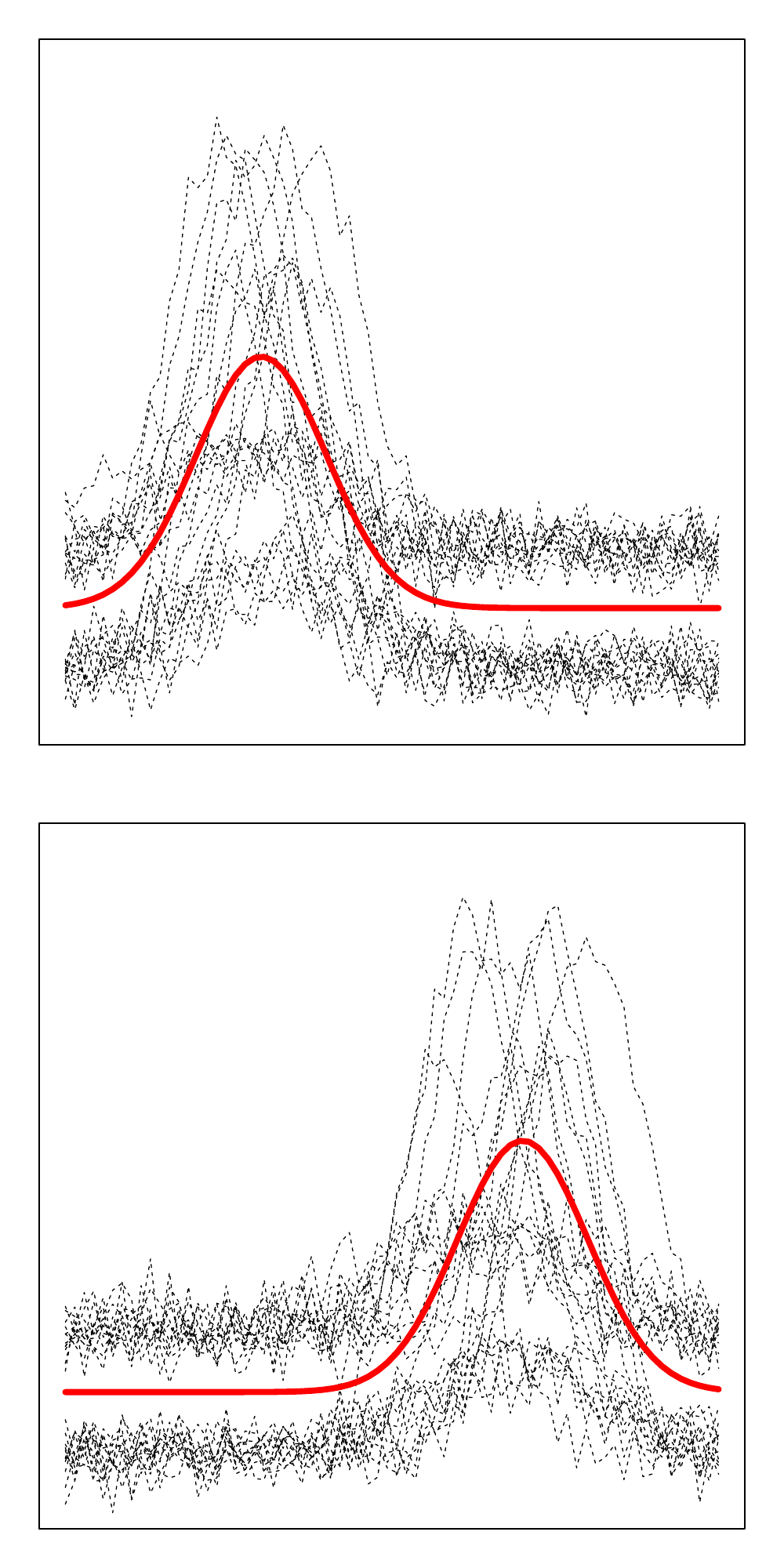}
\caption{Pairs of plots in each column represent the two-cluster configurations arising from switching off, from left to right $\alpha_{ij,1}$, $\alpha_{ij,2},$ $\alpha_{ij,3}$. In the names of the models, as used in the rest of the paper, \texttt{T} indicates a switched on random effect while \texttt{F} a switched off one.}
\label{fig:FT}
\end{figure}

In fact, further flexibility can be naturally introduced within the model by ``switching off'' one or more random effects, depending on subject-matter considerations and on the concept of cluster one has in mind. If there are reasons to support that similar time evolutions associated to different intensities are, in fact, expression of different clusters, it makes sense to switch off the random intercept $\alpha_{ij,1}$. In the example illustrated in Figure \ref{fig:curve_coclust} this ideally leads to a two-clusters structure (Figure \ref{fig:FT}, left panels). Similarly, switching off the random effect $\alpha_{ij,3}$ would lead to blocks characterized by a shifted time evolution (Figure \ref{fig:FT}, right panels), while removing $\alpha_{ij,2}$ determines different blocks varying for a scale factor (Figure \ref{fig:FT}, middle panels). 

%%%%%%%%%%%%%%%%%%%%%%%%%%%%%%%%%%%%%%%%%%%%%%%%%%%%%%%%%%%%%%%%%

\subsection{Model estimation}\label{sec:chcharles_modelest}
To estimate the LBM several approaches have been proposed, as for example Bayesian \citep{wyse2012block}, greedy search \citep{wyse2017inferring} and likelihood-based ones \citep{govaert2008block}. In this work we focus on the latter class of methods. In principle, the estimation strategy would aim to maximize the log-likelihood $\ell(\Theta) = \log p(\mathcal{X}; \Theta)$ with $p(\mathcal{X}; \theta)$ defined as in (\ref{eq:LBM}); nonetheless, the missing structure of the data makes this maximization impractical. For this reason the \emph{complete data log-likelihood} is usually considered as the objective function to optimize, being defined as
\begin{equation}\label{eq:compl_loglik_LBM}
\ell_c(\Theta,\mathbf{z},\mathbf{w}) = \sum_{ik} z_{ik}\log\pi_k + \sum_{jl}w_{jl}\log\rho_l + \sum_{ijkl}z_{ik}w_{jl}\log p(x_{ij}; \theta_{kl})
\end{equation}
where the first two terms account for the proportions of row and column clusters while the third one depends on the probability density function of each block.

As a general solution, to maximize (\ref{eq:compl_loglik_LBM}) and obtain an estimate of $\hat{\Theta}$ when dealing with situations where latent variables are involved, one would instinctively resort to the Expectation-Maximization algorithm \citep[EM,][]{dempster1977maximum}. The basic idea underlying the EM algorithm consists in finding a lower bound of the log-likelihood and optimizing it via an iterative scheme in order to create a converging series of $\hat{\Theta}^{(h)}$ \citep[see][for more details about the convergence properties of the algorithm]{wu1983convergence}. In the co-clustering framework, this lower bound can be easily exhibited by rewriting the log-likelihood as follows
$$
\ell(\Theta) = \mathcal{L}(q;\Theta) +  \zeta 
$$
where $\mathcal{L}(q; \Theta) = \sum_{{\bf z},{\bf w}} q({\bf z},{\bf w})\log(p(\mathcal{X},{\bf z},{\bf w}| \theta)/q({\bf z},{\bf w}))$ with $q({\bf z},{\bf w})$ being a generic probability mass function on the support of $({\bf z},{\bf w})$  while $\zeta$ is a positive constant not depending on $\Theta$. 

The E step of the algorithm maximizes the lower bound $\mathcal{L}$ over $q$ for a given value of $\Theta$. Straightforward calculations show that $\mathcal{L}$ is maximized for $q^{*}({\bf z},{\bf w})=p({\bf z},{\bf w}|\mathcal{X},\theta)$. Unfortunately, in a co-clustering scenario, the joint posterior distribution $p({\bf z},{\bf w}|\mathcal{X},\Theta)$ is not tractable, as it involves terms that cannot be factorized as it conversely happens in a standard mixture model framework. As a consequence, several modifications have been explored, searching for viable solutions when performing the E step  \citep[see][for a more detailed tractation]{govaert2013co}; examples are the \emph{Classification EM} (CEM) and the \emph{Variational EM} (VEM). Here we propose to make use of a Gibbs sampler within the E step to approximate the posterior distribution $p({\bf z},{\bf w}|\mathcal{X},\Theta)$. This results in a stochastic version of the EM algorithm, which will be called SEM-Gibbs in the following.  Given an initial column partition ${\bf w^{(0)}}$ and an initial value for the parameters $\Theta^{(0)}$, at the $h$\emph{-th} iteration the  algorithm proceeds as follows 
\begin{itemize}
	\item SE step: $q^{*}({\bf z},{\bf w})\simeq p({\bf z},{\bf w}|\mathcal{X},\Theta^{(h-1)})$ is approximated with a Gibbs sampler. The Gibbs sampler consists in sampling alternatively ${\bf z}$ and ${\bf w}$ from their conditional distributions a certain number of times before to retain new values for ${\bf z}^{(h)}$ and ${\bf w}^{(h)}$,
	\item M step: $\mathcal{L}(q^{*}({\bf z},{\bf w}),\Theta^{(h-1)})$ is then maximized over $\Theta$, where
	\begin{align*}
	\mathcal{L}(q^{*}({\bf z},{\bf w}),\Theta^{(h-1)}) & \simeq \sum_{z,w}p({\bf z},{\bf w}|\mathcal{X},\Theta^{(h-1)})\log(p(\mathcal{X},{\bf z},{\bf w}|\Theta)/p({\bf z},{\bf w}|\mathcal{X},\Theta^{(h-1)}))\\
		& \simeq E[\ell_c(\Theta, {\bf z}^{(h)}, {\bf w}^{(h)})|\Theta^{(h-1)}]+\xi,
	\end{align*}
	$\xi$ not depending on $\Theta$. This step therefore reduces to the maximization of the conditional expectation of the \emph{complete data log-likelihood} (\ref{eq:compl_loglik_LBM}) given $\bf{z}^{(h)}$ and ${\bf w}^{(h)}$. 
\end{itemize}

In the proposed framework, due to the presence of the random effects, some additional challenges have to be faced. In fact, the maximization of the conditional expectation of (\ref{eq:compl_loglik_LBM}) associated to model (\ref{eq:simmodel_cluster}) requires a cumbersome multidimensional integration in order to compute the marginal density defined as 
\begin{eqnarray}\label{eq:marglikelihood_raneff}
p(x_{ij};\theta_{kl}) = \int p(x_{ij}|\alpha_{ij}^{kl};\theta_{kl})p(\alpha_{ij}^{kl};\theta_{kl})\, d\alpha_{ij}^{kl} \; . 
\end{eqnarray} 
Note that, with a slight abuse of notation, we suppress here the dependency on the time $t$, i.e. $x_{ij}$ has to be intended as $x_{ij}({\bf t_i})$. In the SE step, on the other hand, the evaluation of (\ref{eq:marglikelihood_raneff}) is needed for all the possible configurations of $\{z_i\}_{i=1,\dots,n}$ and $\{w_{j}\}_{j=1,\dots,d}$. These quantities are straightforwardly obtained when considering the SEM-Gibbs to estimate models without any random effect involved, while their computation is more troublesome in our scenario.    

For these reasons, we propose a modification of the SEM-Gibbs algorithm, called \emph{Marginalized SEM-Gibbs} (M-SEM), where an additional \emph{Marginalization step} is introduced to properly account for the random effects.
Given an initial value for the parameters $\Theta^{(0)}$ and an initial column partition $\mathbf{w}^{(0)}$, the $h$\emph{-th} iteration of the M-SEM algorithm alternates the following steps: 
\begin{itemize}
  \item \textbf{Marginalization step}: The single cell contributions in (\ref{eq:marglikelihood_raneff}) to the \emph{complete data log-likelihood} are computed by means of a Monte Carlo integration scheme as follows 
  %the third term of the \emph{complete data log-likelihood} (\ref{eq:compl_loglik_LBM}) is obtained by means of the so called LME approximation proposed in \citet{lindstrom1990nonlinear} alternating between a penalized nonlinear least squares (PNLS) step and a linear mixed effects one. Moreover, in order to perform the SE step of the algorithm, the single contribution of each matrix cell to the. Therefore (\ref{eq:marglikelihood_raneff}) is computed by means of a Monte Carlo integration scheme as follows 
  \begin{eqnarray}
    p(x_{ij};\theta_{kl}^{(q)}) \simeq \frac{1}{M} \sum_{m=1}^M p(x_{ij} ; \alpha_{ij}^{kl,(m)}, \theta_{kl}^{(h-1)}) \; 
  \end{eqnarray}
  for $i=1,\dots,n$, $j=1,\dots,d$, $k=1,\dots,K$ and $l=1,\dots,L$ and being $M$ the number of Monte Carlo samples. The values of the vectors $\alpha_{ij}^{kl,(1)},\dots,\alpha_{ij}^{kl,(M)}$ are drawn from a Gaussian distribution $\mathcal{N}_3(\mu_{kl}^{\alpha,(h-1)},\Sigma_{kl}^{\alpha,(h-1)})$; note that this choice amounts to a random version of the \emph{Gaussian quadrature rule} \cite[see e.g.][]{pinheiro2006mixed};

  \item \textbf{SE step}: $p({\bf z},{\bf w}|\mathcal{X},\Theta^{(h-1)})$ is approximated by repeating, for a number of iterations, the following Gibbs sampling steps
  \begin{enumerate}
    \item generate the row partition $z_i^{(h)}=(z_{i1}^{(h)},\dots,z_{iK}^{(h)}), \; i=1,\dots,n$ according to a multinomial distribution $z_i^{(h)}\sim \mathcal{M}(1,\tilde{z}_{i1},\dots,\tilde{z}_{iK})$, with
    \begin{eqnarray*}
      \tilde{z}_{ik} &=& p(z_{ik}=1 | \mathcal{X},\mathbf{w}^{(h-1)};\Theta^{(h-1)}) \\
      &=& \frac{\pi_k^{(h-1)}p_k(\mathbf{x}_i | \*w^{(h-1)}; \Theta^{(h-1)})}{\sum_{k'}\pi_{k'}^{(h-1)}p_{k'}(\mathbf{x}_i | \*w^{(h-1)}; \Theta^{(h-1)})} \; ,
    \end{eqnarray*}
    for $k=1,\dots,K$, $\mathbf{x}_i = \{x_{ij}\}_{1\le j\le d}$ and $p_k(\mathbf{x}_i | \mathbf{w}^{(h-1)}; \Theta^{(h-1)}) = \prod_{jl} p(x_{ij}; \theta_{kl}^{(h-1)})^{w_{jl}^{(h-1)}}$. 
    \item generate the column partition $w_j^{(h)}=(w_{j1}^{(h)},\dots,w_{jL}^{(h)}), \; j=1,\dots,d$ according to a multinomial distribution $w_j^{(h)}\sim \mathcal{M}(1,\tilde{w}_{j1},\dots,\tilde{w}_{jL})$, with
    \begin{eqnarray*}
      \tilde{w}_{jl} &=& p(w_{jl}=1 | \mathcal{X}, \mathbf{z}^{(h)}; \Theta^{(h-1)}) \\
      &=& \frac{\rho_l^{(q)}p_l(\mathbf{x}_j | \mathbf{z}^{(h)}; \Theta^{(h-1)})}{\sum_{l'}\rho_{l'}^{(h-1)}p_{l'}(\mathbf{x}_j | \mathbf{z}^{(h)}; \Theta^{(h-1)})} \; , 
    \end{eqnarray*}
    for $l=1,\dots,L$, $\mathbf{x}_j = \{x_{ij}\}_{1 \le i\le n}$ and $p_l(\mathbf{x}_j | \mathbf{z}^{(h)}; \Theta^{(h-1)}) = \prod_{ik} p(x_{ij}; \Theta_{kl}^{(h-1)})^{z_{ik}^{(h)}}$.
  \end{enumerate}
  \item \textbf{M step}: Estimate $\Theta^{(h)}$ 
  %conditionally on $\mathbf{z}^{(h)}$ and $\mathbf{w}^{(h)}$ 
  by maximizing $E[\ell_c(\Theta, {\bf z}^{(h)}, {\bf w}^{(h)})|\Theta^{(h-1)}]$.\\ Mixture proportions are updated as $\pi_k^{(h)} = \frac{1}{n}\sum_{i}z_{ik}^{(h)}$ and $\rho_l^{(h)}=\frac{1}{d}\sum_j w_{jl}^{(h)}$. The  estimate of $\theta_{kl}=(\mu_{kl}^\alpha,\Sigma_{kl}^{\alpha},\sigma^2_{\epsilon,kl},\beta_{kl})$ is obtained by exploiting the \emph{non-linear mixed effect model} specification in (\ref{eq:simmodel_cluster}) and considering the approximate maximum likelihood formulation proposed in \citet{lindstrom1990nonlinear}; the variance and the mean components are estimated by approximating and maximizing the marginal density of the latter near the mode of the posterior distribution of the random effects. Conditional or shrinkage estimates are then used for the estimation of the random effects. 
\end{itemize}

The M-SEM algorithm is run for a certain number of iterations until a convergence criterion on the \emph{complete data log-likelihood} is met. Since a burn-in period is considered, the final estimate for $\Theta$, denoted as $\hat{\Theta}$, is given by the mean of the sample distribution. A sample of $(\mathbf{z},\mathbf{w})$ is then generated according to the SE step as illustrated above with $\Theta=\hat{\Theta}$. The final block-partition $(\hat{\mathbf{z}},\hat{\mathbf{w}})$ is then obtained as the mode of their sample distribution. 

\subsection{Model selection}
%% EE: I suggest to use "dimensionality selection" instead of "model selection" whereever you are only talking about selecting $K$.

The choice of the number of groups is considered here
% recasted 
as a model selection problem. Operationally several models, corresponding to different combinations of $K$ and $L$ and, in our case, to different configurations of the random effects, are estimated and the best one is selected according to an information criterion. Note that the model selection step is more troublesome in this setting with respect to a standard clustering one, since we need to select not only the number of row clusters $K$ but also the number of column ones $L$.  Standard choices, such as the AIC and the BIC, are not directly available in the co-clustering framework where, as noted by \citet{keribin2015estimation}, the computation of the likelihood of the LBM is challenging, even when the parameters are properly estimated. Therefore, as a viable alternative, we consider an approximated version of the ICL \citep{biernacki2000assessing} that, relying on the \emph{complete data log-likelihood}, does not suffer from the same issues, and which reads as follows
\begin{eqnarray}\label{eq:ICL_cocl}
\text{ICL} = \ell_c(\hat{\Theta}, \hat{z}, \hat{w}) - \frac{K-1}{2}\log n - \frac{L-1}{2}\log d - \frac{KL\nu}{2}\log nd \; , 
\end{eqnarray}
where $\nu$ denotes number of specific parameters for each block while $\ell_c(\hat{\Theta}, \hat{z}, \hat{w})$ is defined as in \eqref{eq:compl_loglik_LBM} with $\Theta$, $z$ and $w$ being replaced by their estimates.
%\begin{eqnarray}\label{eq:compl_densfunct}
%\ell_c(\hat{\Theta}, \hat{z}, \hat{w}) = \sum_{ik} \hat{z}_{ik}\log\hat{\alpha}_k + \sum_{jl}\hat{w}_{jl}\log \hat{\rho}_l + \sum_{ijkl} \hat{z}_{ik}\hat{w}_{jl}\log p(x_{ij}; \hat{\theta}_{kl}) \; .
%\end{eqnarray}
The model associated with the highest value of the ICL is then selected. 

%Even if this criterion has shown good performances in many co-clustering applications, 
Even if the use of this criterion is a well-established practice in co-clustering applications, \citet{keribin2015estimation} noted that its consistency has not been proved yet to estimate the number of blocks of a LBM. Additionally, \citet{nagin2009group} and \citet{corneli2020bayesian} point out a bias of the ICL towards overestimation of the number of clusters in the longitudinal context. The validity of the ICL could be additionally undermined by the presence of random effects. As noted by \citet{delattre2014note}, standard information criteria have unclear definitions in a mixed effect model framework, since the definition of the actual sample size is not trivial. Given that, common asymptotic approximations are not valid anymore. Even if a proper exploration of the problem from a co-clustering perspective is still missing, we believe that the mentioned issues might have an impact also on the derivation of the criterion in (\ref{eq:ICL_cocl}). The development of valid model selection tools for LBM when random effects are involved is out of scope of this work, therefore, operationally, we consider the ICL. Nonetheless, the analyses in Section \ref{sec:chcharles_numexample} have to be interpreted with full awareness of the limitations described above. 
% these considerations. 

%%%%%%%%%%%%%%%%%%%%%%%%%%%%%%%%%%%%%%%%%%%%%%%%%%%%%%%%%%%%%%%%%

\subsection{Remarks}\label{sec:chcharles_discussion}
The model introduced so far inherits the advantages of both the building ingredients, namely the LBM and the SIM, it embeds. 
%% EE: What are the two "ingredients"? More than two are mentioned in this paragraph which is confusing. 
Thanks to the local independence assumption of the LBM, it allows handling multivariate, possibly high-dimensional complex data structures in a relatively parsimonious way. Differences among the subjects are captured by the random effects, while curve summaries can be expressed as a functional of the mean shape curve. Additionally, resorting to a smoother when modeling the mean shape function, it allows for a flexible handling of functional data whereas the presence of random effects make the model effective also in a longitudinal setting. Finally, clustering is pursued directly on the observed curves, without resorting to intermediate transformation steps, as it is done in \citet{bouveyron2018functional}, where clustering is performed on the basis expansion coefficients used to transform the original data. 

The attractive features of the model go hand in hand with some difficulties that require caution --namely, likelihood multimodality, the associated convergence issues, and the curse of flexibility-- that we discuss below in turn.  
% These reasons of attractiveness should not distract from the caution required by some aspects of the proposed method, discussed in the following.
\begin{itemize}
\item \emph{Initialization} The M-SEM algorithm encloses different numerical steps which require the suitable specification of starting values. 
First, the convergence of EM-type algorithms towards a global maximum is not guaranteed; as a consequence they are known to be sensitive to the initialization with a proper one being crucial to avoid local solutions. Assuming $K$ and $L$ to be known, the M-SEM algorithm requires starting values for $\mathbf{z}$ and $\mathbf{w}$ in order to implement the first M step. A standard strategy resorts to multiple random initializations: the row and column partitions are sampled independently from multinomial distributions with uniform weights and the one eventually leading to the highest value of the \emph{complete data log-likelihood} is retained. An alternative approach, possibly accelerating the convergence, is given by a k-means initialization, where two k-means algorithms are independently run for the rows and the columns of $\mathcal{X}$ and the M-SEM algorithm is initialized with the obtained partitions. It has been pointed out \citep[see e.g.][]{govaert2013co} that the SEM-Gibbs, being a stochastic algorithm, can attenuate in practice the impact of the initialization on the resulting estimates.  Finally, note that a further initialization is required, to estimate the nonlinear mean shape function within the M step. 

\item \emph{Convergence and other numerical problems}. Although the benefits of including random effects in the considered framework 
are undeniable, parameters estimation is known not to be straightforward in mixed effect models,  especially in the nonlinear setting \citep[see, e.g.][]{harring2016comparison}. As noted above the nonlinear dependence of the conditional mean of the response on the random effects requires 
multidimensional integration to derive the
marginal distribution of the data. While several methods have been proposed to compute the integral, convergence issues are often encountered. 
In such situations, some strategies can be employed to help with convergence of the estimation algorithm. Examples are to try different sets of starting values, to scale the data prior to the modeling step, or to simplify the structure of the model (e.g. by reducing the number of knots of the B-splines). Addressing these issues often results in considerable computational times even when convergence is eventually achieved. Depending on the specific data at hand, it is also possible to consider alternative mean shape formulations, such as polynomial functions, which result in easier estimation procedures. Lastly note that, if available, prior knowledge about the time evolution of the observed phenomenon may be incorporated in the models as it could introduce some constraints possibly simplifying the estimation process \citep[see e.g.][]{telesca2012modeling}.

\item \emph{Curse of flexibility}. Including random effects for both phase and amplitude shifts and scale transformations might allow for a virtually excellent fitting of various arbitrarily shaped curve. This flexibility, albeit desirable, sometimes achieve excessive extents, possibly leading to estimation troubles. This is especially true in a clustering framework, where data are expected to exhibit a remarkable heterogeneity. 
From a practical point of view, our experience suggests that the estimation of the parameters $\alpha_{ij,2}$ turns out to be the most troublesome, sometimes leading to convergence issues and instability in the resulting estimates.

\end{itemize}

%%%%%%%%%%%%%%%%%%%%%%%%%%%%%%%%%%%%%%%%%%%%%%%%%%%%%%%%%%%%%%%%%

\section{Numerical experiments}\label{sec:chcharles_numexample}

Before moving forward to the interest of the proposed approach on real-world situations, this section presents some numerical experiments illustrating its main features on simulated data. 

\subsection{Synthetic data}\label{sec:chcharles_simulation}

This section examines the main features of the proposed approach on some synthetic data. The aim of the simulation study is twofold. The first and primary goal of the analyses consists in exploring the capability of the proposed method to properly partition the data into blocks, also in comparison with some competitors such as the one proposed by \citet{bouveyron2018functional} (\texttt{funLBM} in the following) and a double k-means approach, where row and column partitions are obtained separately and subsequently merged to produce blocks. As for the second aim of the simulations, we evaluate the performances of the ICL in the developed framework to select both the number $(K,L)$ of blocks and the random effect configuration.
%In order to isolate the different sources of uncertainty, the two tasks are pursued separately. 
%Our proposal is compared, in terms of co-clustering quality, with some competitors such as the one proposed by \citet{bouveyron2018functional} (\texttt{funLBM} in the following) as well as a double k-means approach, where row and column partitions are obtained separately and subsequently merged to produce blocks. 
%The distinct scopes of the simulations find their rationales in the necessity to isolate the different sources of uncertainty. Coherently with the discussion in Section \ref{sec:chcharles_discussion}, the performances of the proposed procedure might be influenced by the model selection strategy adopted as well as by some subjective choices that have to be made in the model specification step; by distinguishing among different objectives we aim to obtain a proper evaluation of the merits and the drawbacks of the considered co-clustering strategy. \\ 

All the analyes have been conducted in the R environment \citep{rcore} with the aid of \texttt{nlme} package \citep{nlmepack}, used to estimate the parameters in the M-step, and the \texttt{splines} package, used to handle the B-spline involved in the specification of the common shape function. The code implementing the proposed procedure is available upon request.

\begin{figure}[!t]
  \centering
  \includegraphics[width=14cm,height=8.5cm]{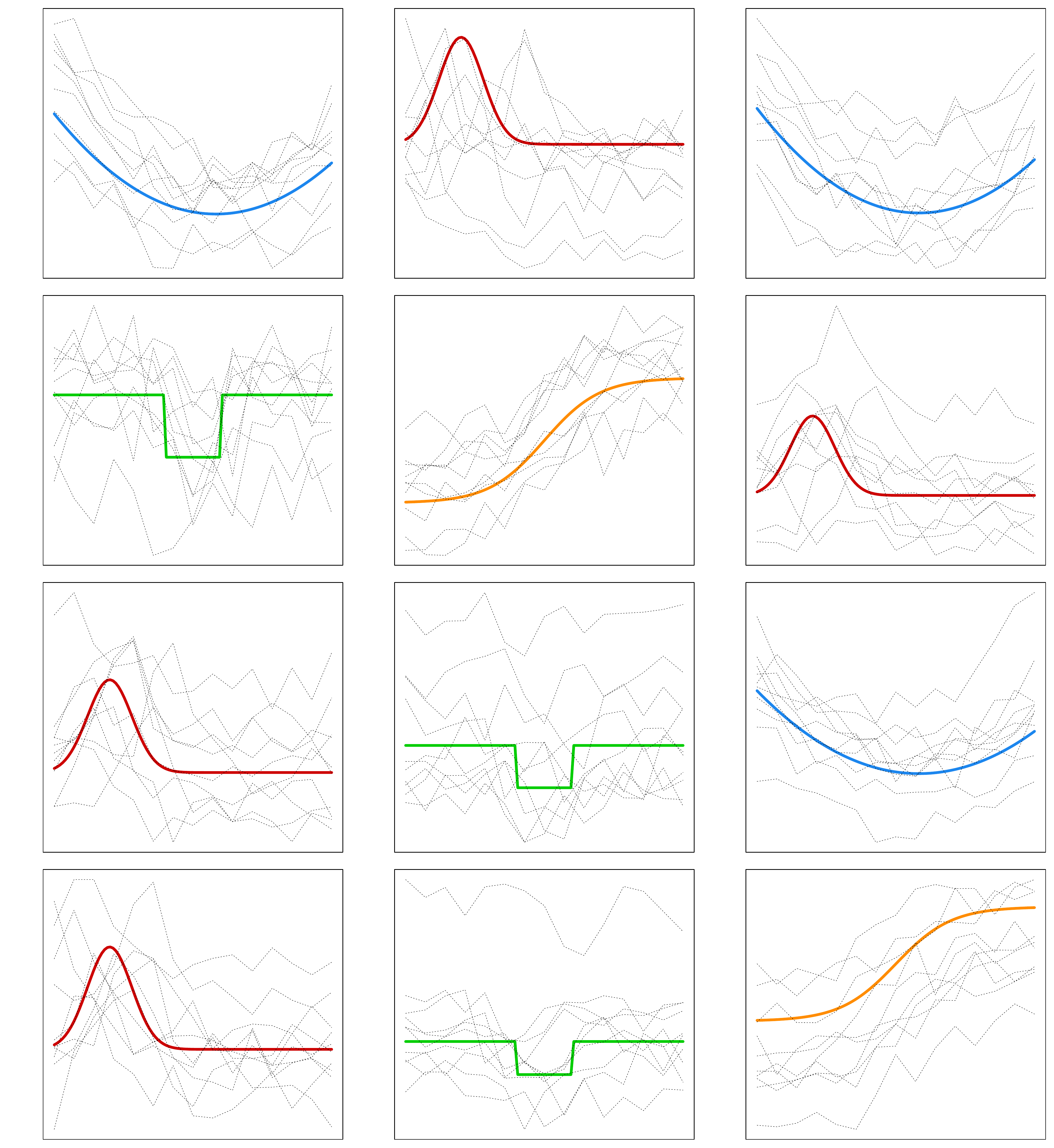}
  \caption{Subsample of simulated curves (black dashed lines) with over-imposed block specific mean shape curves (colored continue lines) employed in the numerical study}
  \label{fig:meancurves}
\end{figure}

The examined simulation setup is defined as follows. We generated $B=100$ Monte Carlo samples of curves according to the general specification \eqref{eq:simmodel_cluster}, with block-specific mean shape function $m_{kl}(\cdot)$ and both the parameters involved in the error term and the ones describing the random effects distribution being constant across the blocks. In fact, in the light of the considerations made in Section \ref{sec:chcharles_discussion}, the random scale parameter is switched off in the data generative mechanism, i.e. $\alpha_{ij,2}$ is constrained to be degenerate in zero. The number of row and column clusters has been fixed to $K_{\text{true}} = 4$ and $L_{\text{true}} = 3$. The mean shape functions $m_{kl}(\cdot)$ are chosen among four different curves, namely $m_{11}=m_{13}=m_{33}=m_1$, $m_{12}=m_{32}=m_{31}=m_{41}=m_2$, $m_{21}=m_{32}=m_{42}=m_3$ and $m_{22}=m_{43}=m_4$, as illustrated in Figure \ref{fig:meancurves} with different color lines, and specified as follows: 
\begin{eqnarray*}
m_1(t) &\propto& 6t^2 - 7t + 1 \hspace{3cm}  m_2(t) \propto \phi(t; 0.2,0.008) \\
m_3(t) &\propto& 0.75-0.8\mathbbm{1}_{\{t \in (0.4,0.6)\}} \hspace{1.4cm} m_4(t) \propto \frac{1}{(1 + \exp(-10t + 5))} 
\end{eqnarray*} 
The other parameters involved have been set to $\sigma_{\epsilon,kl}=0.3$, $\mu_{kl}^\alpha = (0,0,0)$ and $\Sigma_{kl}^\alpha = \text{diag}(1,0,0.1)$ $\forall k=1,\dots,K_{\text{true}}, l=1,\dots,L_{\text{true}}$. Three different scenarios are considered with generated curves consisting of $T = 15$ equi-spaced observations ranging in $[0,1]$. As a first baseline scenario, we set the number of rows to $n = 100$ and the number of columns to $d = 20$. The other scenarios are considered in order to obtain insights and indications on the performances of the proposed method when dealing with larger matrices. Coherently in the second scenario $n = 500$ and $d = 20$ while in the third one $n = 100$ and $d = 50$ thus increasing respectively the number of samples and features.

As for the first goal of the simulation study, we explore the performances of our proposal in terms of the Co-clustering Adjusted Rand Index \citep[CARI,][]{robert2020comparing}. This criterion generalizes the Adjusted Rand Index  \citep{hubert1985comparing} to the co-clustering framework, and takes the value 1 when the blocks partitions perfectly agree up to a permutation. 
In order to have a fair comparison with the double \emph{k-means} approach, for which selecting the number of blocks does not admit an obvious solution, and to separate the uncertainty due to model selection from the one due to cluster detection, models are compared by considering the number of blocks as known and equal to $(K_{\text{true}}, L_{\text{true}})$. Consistently, we estimate our model only for the \texttt{TFT} random effects configuration, being the one generating the data. This is made possible by constraining mean and variance estimates of the scale-random effect to be equal to zero. 

Results are reported in Table \ref{tab:charles_simulation1}. 
The proposed method claims excellent performances in all the considered settings, with results notably featured by a very limited variability. No clear-cut indications arise from the comparison with \texttt{funLBM} as the two methodologies lead to comparable results. Conversely, the use of an approach which is not specifically conceived for co-clustering, like the the double \emph{k-means}, leads to a strong degradation of the quality of the partitions. Unlike the results obtained with our method and with \texttt{funLBM}, being insensitive to changes in the dimensions of the data, the double \emph{k-means} behaves better with an increased number of variables or observations. 
%A first indication that emerges is concerned with the performances of the double \emph{k-means} approach. In fact, it appears clear that the use of an approach which is not specifically conceived for co-clustering leads to a strong degradation of the quality of the partitions. Contrary to the results obtained with our method and with \texttt{funLBM}, being insensitive to changes in the dimensions of the data, the double \emph{k-means} behaves better in those scenarios with an increased number of variables or observations. The comparison between the proposed approach and \texttt{funLBM} shows less clear-cut indications as the two methodologies lead to comparable results in all the three scenarios considered. Therefore our proposal leads to similar conclusions with respect to \texttt{funLBM} while, at the same time, providing a more flexible and interpretable co-clustering mechanism being able to accommodate different notion of groups, as discussed in Section \ref{sec:chcharles_modspec}. 
\begin{table}[t]
\centering
\caption{Mean (and std error) of the CARI computed over the simulated samples in the three scenarios. Partitions are obtained using the proposed approach (\texttt{tdLBM}), \texttt{funLBM} and a double k-means approach.}
\vspace*{0.2cm}
\begin{tabular}{cccc}
  \hline
 & $n = 100, d = 20$ & $n = 100, d = 50$ & $n = 500, d = 20$ \\ 
  \hline
CARI$_{\text{\texttt{tdLBM}}}$ & 0.972 (0.044) &  0.988 (0.051)  &  0.981 (0.020) \\ 
CARI$_{\text{\texttt{funLBM}}}$ & 0.981 (0.066) & 0.986 (0.053) & 0.986 (0.060) \\ 
CARI$_{\text{\texttt{kmeans}}}$ & 0.761 (0.158) & 0.842 (0.182) & 0.809 (0.169) \\ 
   \hline
\end{tabular}
\label{tab:charles_simulation1}
\end{table}

\begin{table}[b]
\centering
  \caption{Rate of selection of $(K,L)$ configurations for the different scenarios considered obtained using the proposed approach.}
  \vspace*{0.3cm}
\begin{tabular}{p{4cm}p{4cm}p{4cm}p{2cm}}
\hline
\hspace{1.15cm} $n=100, d=20$ & \hspace{0.85cm} $n=100, d=50$ & \hspace{0.7cm} $n=500, d=20$&\\
\hline
\multicolumn{3}{l}{\multirow{9}{*}{\includegraphics[height = 4.6cm, width = 13cm]{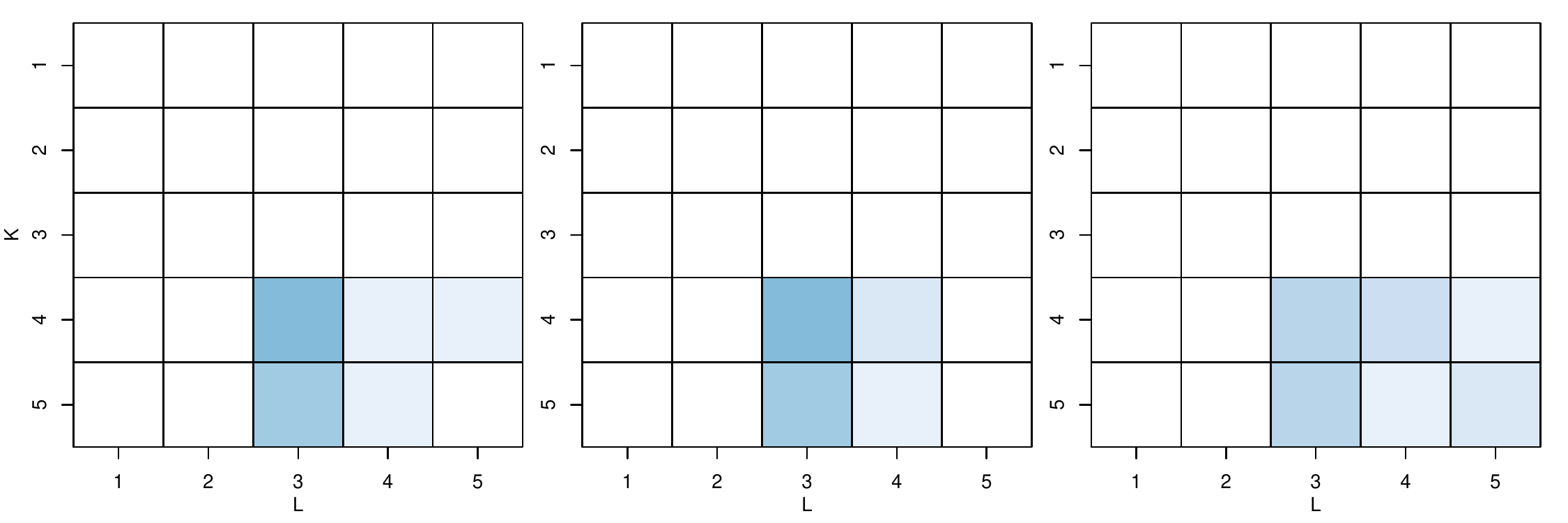}}} & {\multirow{9}{*}{\includegraphics[scale=0.43]{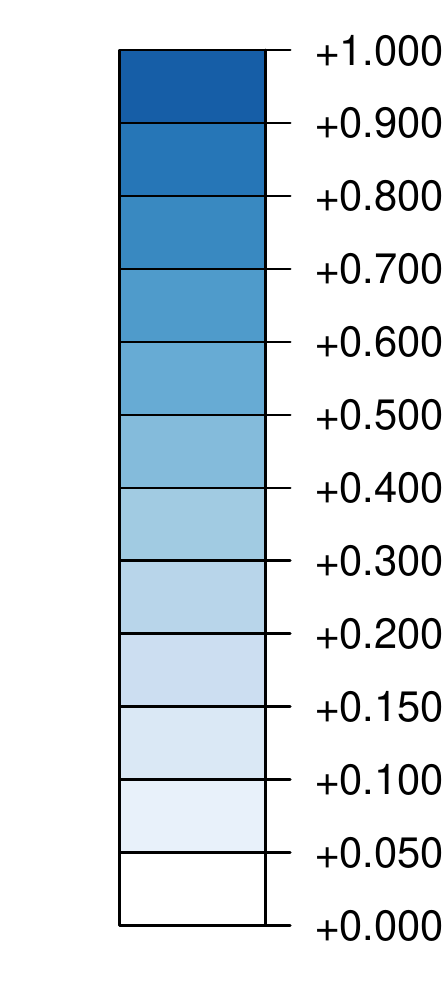}}}\\
%\multirow{12}{*}{\includegraphics[width=.6\textwidth]{Figures/table2.pdf}}&&&\multirow{12}{*}{\includegraphics[scale=0.55]{Figures/output.pdf}}\\
&&&\\
&&&\\
&&&\\
&&&\\
&&&\\
&&&\\
&&&\\
&&&\\
&&&\\
\hline
\end{tabular}
\label{tab:charles_simulation2}
\end{table}

As for the performances of the ICL, Table \ref{tab:charles_simulation2} shows the fractions of samples where the criterion has led to the selection of each of the considered configurations of $(K, L)$, with $K,L = 2,\dots,5$, for models estimated with the proposed method. In all the considered settings, the actual number of co-clusters is the most frequently selected by the ICL criterion. In fact, a non-negligible tendency to favor overparameterized models, especially for larger sample size, is witnessed, consistently with the comments in \citet{corneli2019co}.

Similar considerations may be drawn from the exploration of the performances of the ICL when used to select the random effect configuration (Table \ref{tab:charles_simulation3}). Here, the performances seem to be less influenced by the data dimensionality, and ICL selects the true configuration for the majority of the samples in two scenarios while, in the third one, the true model is selected approximately one out of two samples. Nonetheless, also in this case, a tendency to overestimation is visible, with the \texttt{TTT} configuration frequently selected in all the scenarios. In general, the penalization term in (\ref{eq:ICL_cocl}) seems to be too weak and overall not completely able to account for the presence of random effects. These results, along with the remarks at the end of Section \ref{sec:chcharles_modelest}, provide a suggestion about a possibly fruitful research direction which consists in proposing some suitable adjustments. 

In fact, it is worth noting that when the selection of the number of clusters is the aim, the observed behavior is overall preferable with respect to underestimation since it does not undermine the homogeneity within a block, being the final aim of cluster analysis; this has been confirmed by further analyses suggesting that the additional groups are usually small and arising because of the presence of outliers. As for the random effect configuration, we believe that since the choice impacts  the notion of cluster one aims to identify, it should be driven by subject-matter knowledge rather than by automatic criteria. Additionally, the reported analyses are exploratory in nature, aiming to provide general insights on the characteristics of the proposed approach. To limit computational time required to run the large number of models involved in Tables \ref{tab:charles_simulation2} and \ref{tab:charles_simulation3}, for demonstration of our method, we did not use multiple initializations and we have pre-selected the number of knots for the block-specific mean functions. In practice, we recommend, at a minimum, using multiple starting values and carrying out sensitivity analyses on the number of knots to ensure that substantive conclusions are not affected.

\begin{table}[t]
\centering
  \caption{Rate of selection for each random effects configuration in the considered scenarios. \texttt{T} means that the corresponding random effect is switched on while \texttt{F} means that is switched off. As an example \texttt{FTT} represent a model where $\alpha_{ij,1}$ is constrained to be a random variable with degenerate distribution in zero. Bold cells represents the true data generative model (TFT), blank ones represent percentages equal to zero.}
  \vspace*{0.3cm}
\begin{tabular}{cc|cccccccc}
    \hline
     & & \texttt{FFF} & \texttt{TFF} & \texttt{FTF} & \texttt{FFT} & \texttt{TTF} & \texttt{TFT} & \texttt{FTT} & \texttt{TTT} \\
     \hline 
     & $n = 100, d=20$ & &  &  & & 1\% & {\bf 58\%} & & 41\%\\
     \% of selection 
     & $n = 100, d=50$ &  & 2\% &  &  & 1\% & {\bf 62\%} &  & 35\% \\
   & $n = 500, d=20$ &  & 1\% &  & & 5\% & {\bf 47\%} & & 47\%\\
    \hline
\end{tabular}
\label{tab:charles_simulation3}
\end{table}

%We believe that the reported analyses are somehow exploratory in nature, aiming to provide general insights on the behavior of the proposed approach. Due to some of the complexities mentioned in Section \ref{sec:chcharles_discussion}, a thorough exploration of its performances would be highly time consuming since it would require a proper tuning of some parameters and subjective choices such as the random effects configuration, the number of knots or the smoother to be considered to model the block specific mean shape functions. Some possible further analyses, such as the ones involving alternative choices for the number of observed time points as well as the ones involving the above-mentioned parameters, may improve the understanding of the strengths and weaknesses of the method and constitute interesting developments. 

%%%%%%%%%%%%%%%%%%%%%%%%%%%%%%%%%%%%%%%%%%%%%%%%%%%%%%
\subsection{Applications to real world problems}
\subsubsection{French Pollen Data}\label{sec:chcharles_realdata}
The data we consider in this section are provided by the \emph{R\'eseau National de Surveillance Aérobio-\\logique} (RNSA), the French institute which analyzes the biological particles content of the air and studies their impact on the human health.
RNSA collects data on concentration of pollens and moulds in the air, along with some clinical data, in more than 70 municipalities in France. 

The analyzed dataset contains daily observations of the concentration of 21 pollens for 71 cities in France in 2016. Concentration is measured as the number of pollens detected over a cubic meter of air and carried on by means of some pollen traps located in central urban positions over the roof of buildings, in order to be representative of the trend air quality. 

The aim of the analysis is to identify homogeneous trends in the pollen concentration over the year and across different geographic areas.  
For this reason,  we focus on finding groups of pollens differentiating one from the others for either the period of maximum exhibition or the time span they are present. Consistently with this choice, only models with the y-axis shift parameter $\alpha_{ij,1}$ are estimated (i.e. $\alpha_{ij,2}$ and $\alpha_{ij,3}$ are switched off), for varying number of row and column clusters, and the best one selected via ICL. 
We consider monthly data by averaging the observed daily concentrations over each month. The resulting dataset may be represented as a matrix with $n=71$ rows (cities), $p=21$ columns (pollens) where each entry is a sequence of $T=12$ time-indexed measurements. Moreover, to practically apply our proposed procedure on the data, a preprocessing step has been carried out. We work on a logarithmic scale and, in order to improve the stability of the estimation procedure, the data have been standardized.

\begin{figure}[t]
    \includegraphics[scale=0.2]{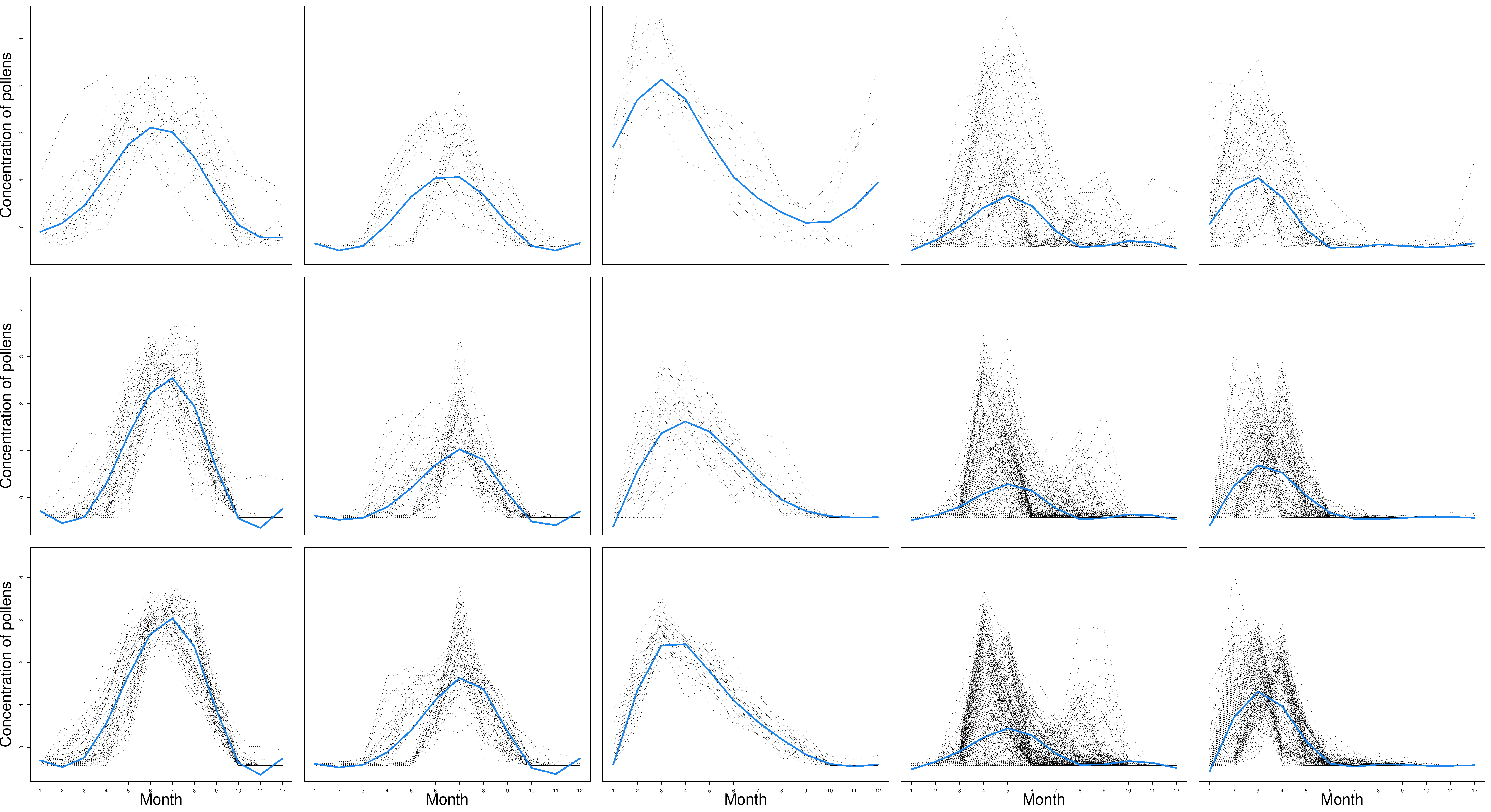}
    \caption{French Pollen Data results. Curves belonging to each single block with superimposed the corresponding block specific mean curve (in light blue).}
    \label{fig:meancurv_pollini}
\end{figure}

Results are graphically displayed in Figure \ref{fig:meancurv_pollini}. The ICL selects a model with $K=3$ row clusters and $L=5$ column ones. % \ref{fig:mappa_francia} 
A first visual inspection of the pollen time evolutions reveals that the procedure is able to discriminate the pollens according to their seasonality. Pollens in the first two column groups are mainly present during the summer, with a difference in the intensity of the concentration.  In the remaining three groups pollens are more active, roughly speaking, during winter and spring months but with a different time persistence and evolution.

Digging deeper substantially in the cluster configuration obtained is beyond the scope of this work and may benefit from insights from experts of botanical and geographical disciplines. Anyway it stands out that column clusters are roughly grouping together trees pollens, distinguishing them from weed and grass ones (right panel of Table \ref{fig:mappa_francia}). Results align with the usually considered typical seasons, with groups of pollens from trees mostly present in winter and spring while the ones from grass spreading in the air mainly during the summer months. With respect to the row partition, displayed in the left panel of Table \ref{fig:mappa_francia}, three clusters have been detected, with one roughly corresponding to the Mediterranean region. The situation, for what it concerns the other two clusters, appears to be more heterogeneous. One of these groups tends to gather cities in the northern region and on the Atlantic coast while the other covers mainly the central and continental part of the country. Again the results appear promising but it may be beneficial a cross analysis with some climate scientists in order to get a more informative and substantiated point of view.

\begin{table}[tb]
\caption{French map with overimposed the points indicating the cities colored according to their row cluster memberships (left) and Pollens organized by the column cluster memberships (right).}\label{fig:mappa_francia}
\begin{center}
\begin{tabular}{c|cl}
 Row groups (Cities) & \multicolumn{2}{c}{Column groups (Pollens)} \\
\hline
\multirow{10}{*}{\vspace{-0.1cm} \includegraphics[scale=0.13]{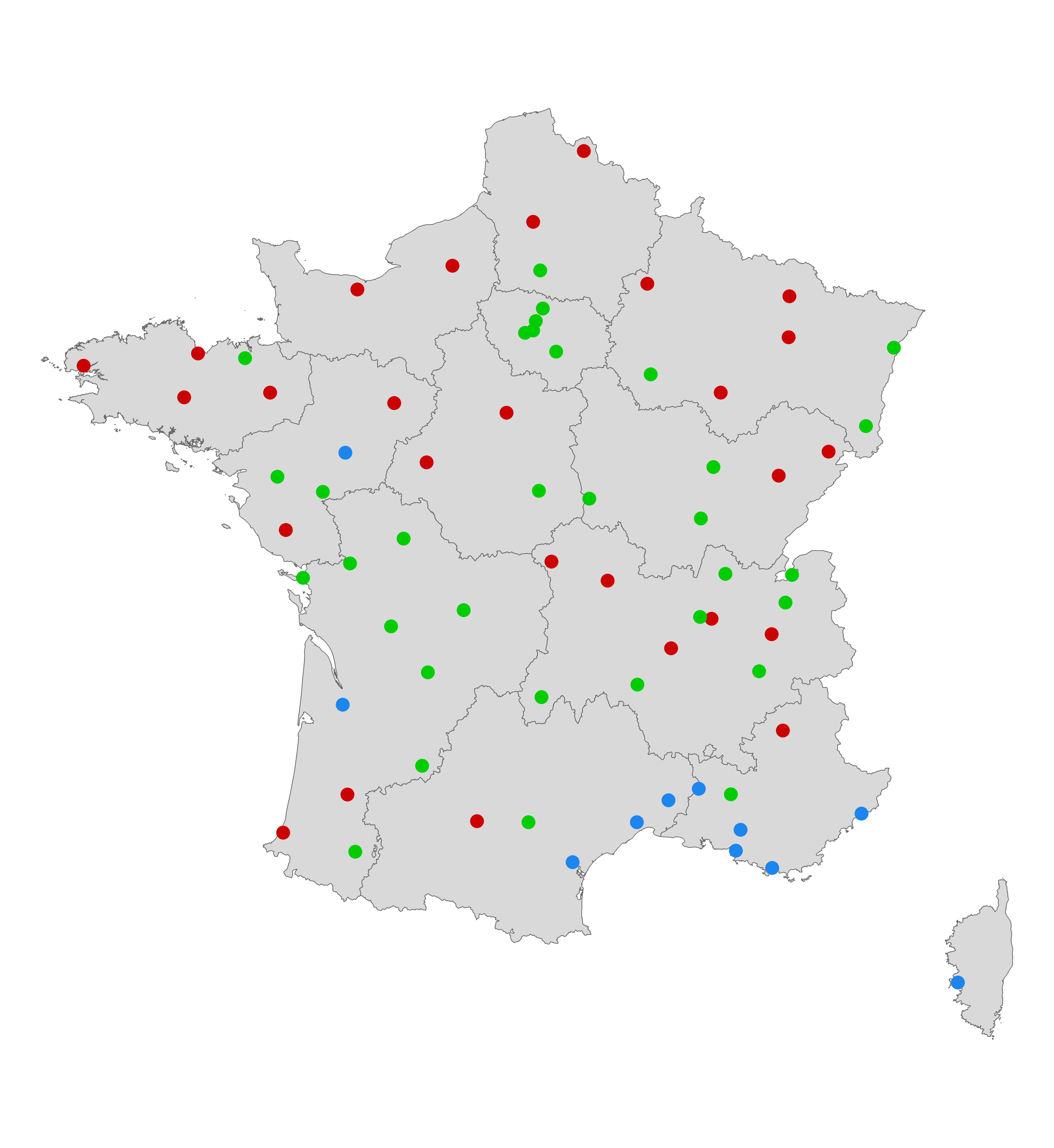}} & & \\
& & \\
& 1 & Gramineae, Urticaceae \\
&  2 & Chestnut, Plantain \\
&  3 & Cypress  \\
&  4 & Ragweed, Mugwort, Birch, Beech, \\ 
&    &  Morus, Olive, Platanus, Oak, Sorrel \\
&    & Linden \\
&  5 & Alder, Hornbeam, Hazel, Ash, \\
&    & Poplar, Willow \\
& & \\
\end{tabular}
\end{center}
\end{table}

\subsubsection{COVID-19 evolution across countries}\label{sec:covid}

At the time of writing this paper, an outbreak of infection with severe acute respiratory syndrome coronavirus 2 (SARS-CoV-2) has severely harmed the whole world. Countries all over the world have undertaken measures to reduce the spread of the virus: quarantine and social distancing practices have been implemented, collective events have been canceled or postponed, business and educational activities have been either interrupted or moved online. 

While the outbreak has led to a global social and economic disruption, its spreading and evolution, also in relation to the aforementioned non pharmaceutical interventions, have not been the same all over the world \citep[see][for an account of this in the first months of the pandemic]{flaxman2020estimating, brauner2021inferring}. With this regard, the goal of the analysis is to evaluate differences and similarities among the countries and for different aspects of the pandemic. 

Being the overall situation still evolving, and given that testing strategies have significantly changed across waves, we refer to the first wave of infection, considering the data from the 1st of March to the 4th of July 2020, in order to guarantee the consistency of the disease metrics used in the co-clustering. Moreover we restrict the analysis to the European countries. Data have been collected by the Oxford COVID-19 Government Response Tracker \citep[OxCGRT, ][]{covid} and originally refer to daily observations 
of the number of confirmed cases and deaths for COVID-19 in each country. We also select two indicators tracking the individual country intervention in response to the pandemic: the Stringency index and the Government response index. Both indicators are recorded on a 0-100 ordinal scale that represents the level of strictness of the policy and accounts for containment and closure policies. The latter indicator also reflects Health system policies such as information campaigns, testing strategies and contact tracing.  

Data have been pre-processed as follows: daily values have been converted into weekly averages in order to reduce the impact of short term fluctuations and the number of time observations. Rates per 1000 inhabitants have been evaluated from the number of confirmed cases and deaths, and the logarithms applied to reduce the data skewness. All the variables have been standardized.  

The  resulting  dataset is a  matrix with $n= 38$ rows (countries), $p$= 4 columns (variables describing the pandemic evolution and containment), observed over a period of $T= 18$ weeks.
%A few countries have been removed from the analysis, due to the presence of missing values, thus resulting in a final sample of size $N=161$ countries. 
Unlike the French Pollen data, here there is no strong reason to favour one random effect configuration over the others. Conversely, different configurations of random effects would entail different ideas of similarity of virus evolution. Thus, while the presence of random effects would allow to cluster together similar trends yet associated to different intensities, speed of evolution and time of onset, switching the random effects off could result in enhancing such differences via the separation of the trends. 
%Switching off all the random effects, for instance, would lead to clusters which can differ in the intensity of the contagion, speed of evolution, and onset of virus spreading. On the other hand, by keeping the random effects on, it would be possible cluster together similar trends, yet associated to different intensities, speed of evolution and time of onset.  %{\bf aggiustare: esiste un modo migliore di dire time of pandemic diffusion? intendo dire l'inizio della diffusione; forse è più corretto dire il concetto al contrario? se tutti i random effect sono accesi, in linea di principio due paesi appartengono allo stesso gruppo se il virus si è diffuso in maniera confrontabile, anche se i contagi sono diversi in termini di intensità o la diffusione è ritardata, o più veloce in uno dei due paesi; inoltre: qui stiamo parlando solo di cluster di riga ma bisognerebbe fare un cenno anche ai cluster di colonna: riflettendo ora mi viene in mente che se forse avessimo usato i nuovi casi e le nuove morti (e non quelle cumulate) un andamento simile (ma shiftato) poteva significare che le misure hanno avuto effetto. Con i casi cumulati invece non lo si può dire, perché le curve sono necessariamente crescenti. Bisogna metterla giù bene, perché il rischio è far arrivare il messaggio -peraltro vero- che ci interessa solo il clustering e non il coclustering}. 

\begin{figure}[bt]
\begin{center}
    \includegraphics[width=.9\textwidth]{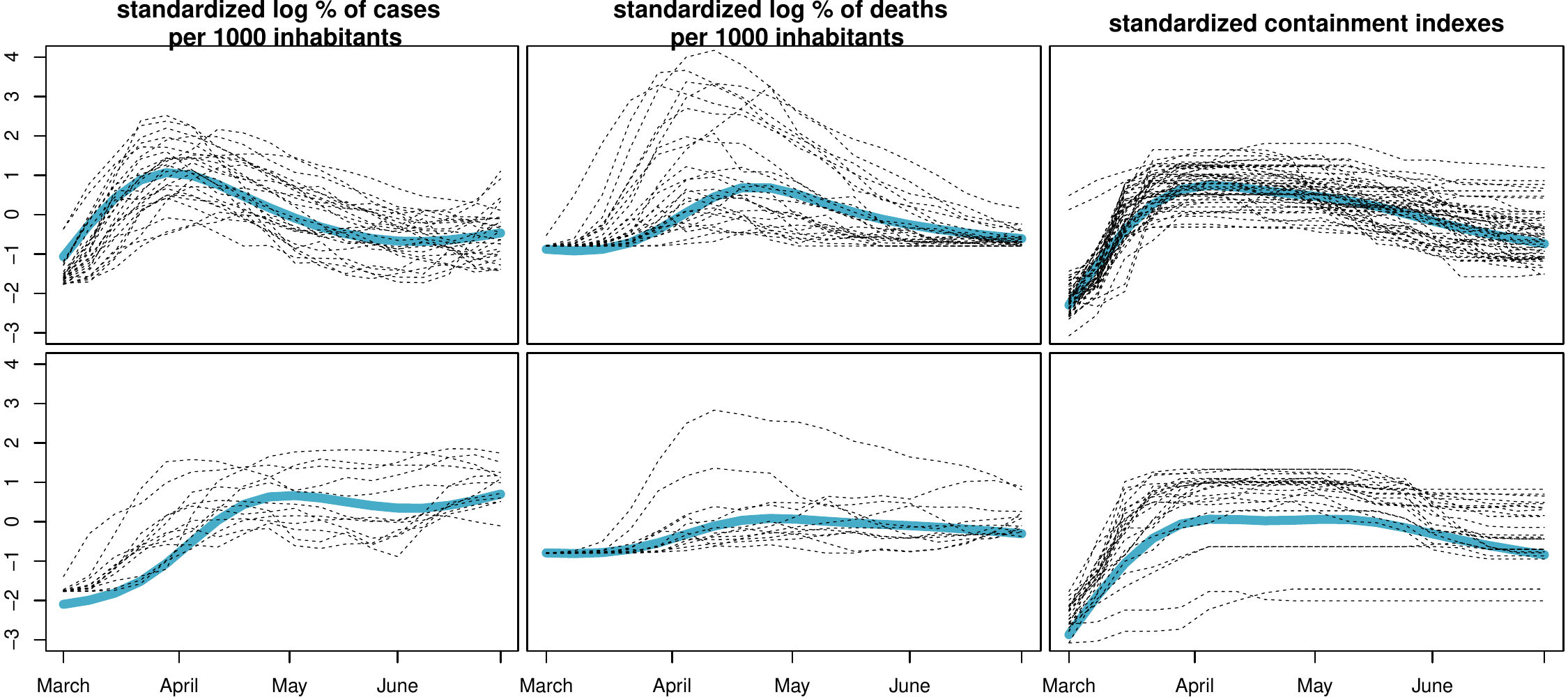}
    \end{center}
    \caption{COVID-19 outgrowth results of the best model, with $K =2,$ $L=3$ and the three random effects on. Curves belonging to each single block with superimposed the associated block specific mean curve (in light blue).}
    \label{fig:best_covid}
\end{figure}

Models have been run for $K = 1, \ldots, 6$ row clusters and $L = 1,2,3$ column clusters, and all the 8 possible configurations of random effects. 
%Estimation has failed for the most parametrized models, whenever data have not shown evidence of splitting in further clusters. %Figure \ref{fig:ICL_covid} displays the ICL associated to the estimated models. Within each of the considered configurations of random effects, there is no remarkable difference between models with a different number of row or column clusters, especially with the most parametrized models in terms of random effects. 
The behaviour of the resulting ICL values supports the remark in Section  \ref{sec:chcharles_simulation}, as the criterion favours highly parameterized models. This holds particularly true with regard to the random effects configuration where the larger the number of random effects switched on, the highest the corresponding ICL. 
%models with none or only one random effect switched on require a larger number of blocks to fit the data adequately. Conversely, 
Thus, models with all the random effects switched on stand out among the others, with a preference for $K=2$ and $L=3$ whose results are displayed in Figure \ref{fig:best_covid}. The obtained partition is easily interpretable: in the column partition, reported on the right panel of Table \ref{fig:best_covid_mappa}, the containment indexes are grouped together into the same cluster whereas the log-rate of positiveness and death are singleton clusters. Consistently with the random effect configuration, row clusters exhibit a different evolution in terms of cases, deaths and undertaken containment measures: 
one cluster gathers countries where the virus has spread earlier and caused more losses; here, more severe control measures have been adopted, whose effect is likely seen in a general decreasing of cases and deaths after achieving a peak. The second row cluster collects countries for which the death toll of the pandemic seems to be more contained. The virus outbreak generally shows a delayed onset and a slower growth, which does not show a steep decline after reaching the peak, although the containment policies remain high for a long period. 
Notably, the row partition is also geographical, with the countries with higher mortality all belonging to the West Europe (see the left panel in Table \ref{fig:best_covid_mappa}).  

\begin{figure}[h]
    \includegraphics[width=.9\textwidth]{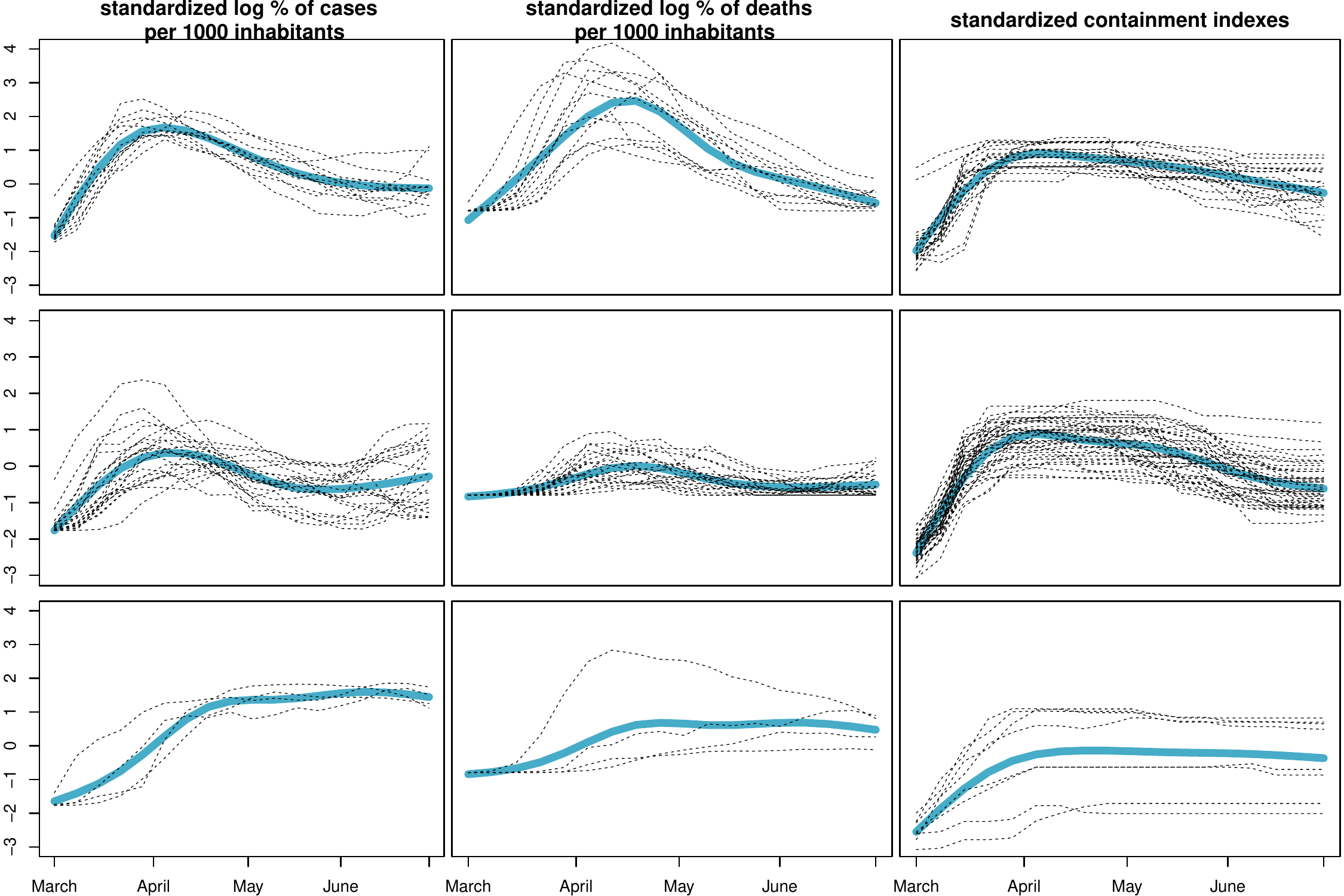}
    \caption{COVID-19 outgrowth results of the model with $K =3,$ $L=3$ and the three random effects off.}
    \label{fig:covid2}
\end{figure}

To properly show the benefits of considering different random effects configurations in terms of notion and interpretation of the clusters, we also illustrate the partition produced by another model estimated having the three random effects switched off (Figure \ref{fig:covid2}). Here we consider $K = L = 3$: the column partition remains unchanged with respect to the best model, with the row partition still separating countries by the severity of the impact, with the third additional cluster having intermediate characteristics. According to this model, two row clusters feature countries with a similar right-skewed bell-shaped trend of cases and similar policies of containment, yet with a notable difference in the virus lethality. Indeed, the effect of switching $\alpha_2$ off is clearly noted in the log-rate of death fitting, with two mean curves having similar shapes but different scales. The additional intermediate cluster, less impacted in terms of death rate, is populated by countries from the central-east Europe. %In fact, al terzo gruppo con pochi morti, confluiscono paesi che prima stavano inentrambi i gruppi, e la spartizione in est-overst europa diventa ancora più evidente {\bf (valutare se meglio mettere 33FTF, sia perché è un po' più coerente con l'interpretazione, sia perché il modello completamente privo di random effect diventa molto diverso da quello specificato nella proposta: l'ho guardato e i modelli mi sembrano coerenti con la spiegazione data, ma forse c'è un fit peggiore)} 
The apparent smaller impact of the first wave of the pandemic on the eastern European countries could be explained by several factors ranging from demographic characteristic and more timely closure policies to a different international mobility pattern. Additionally, other factors such as the general economic and health conditions might have prevented accurate testing and tracking policies, so that the actual spreading of the pandemic might have been underestimated. 

\begin{table}[tb]
\caption{Europe map with countries colored according to their row cluster memberships (left) and variables organized by the column cluster membership (right) for the best ICL model.}
\label{fig:best_covid_mappa}
\begin{center}
\begin{tabular}{p{.56\textwidth}|c p{.34\textwidth}}
 Row groups (Countries) & \multicolumn{2}{p{.4\textwidth}}{Column groups (COVID-19 spreading and containment)} \\
\hline
\multirow{7}{*}{\vspace*{-3cm} \includegraphics[scale=0.43]{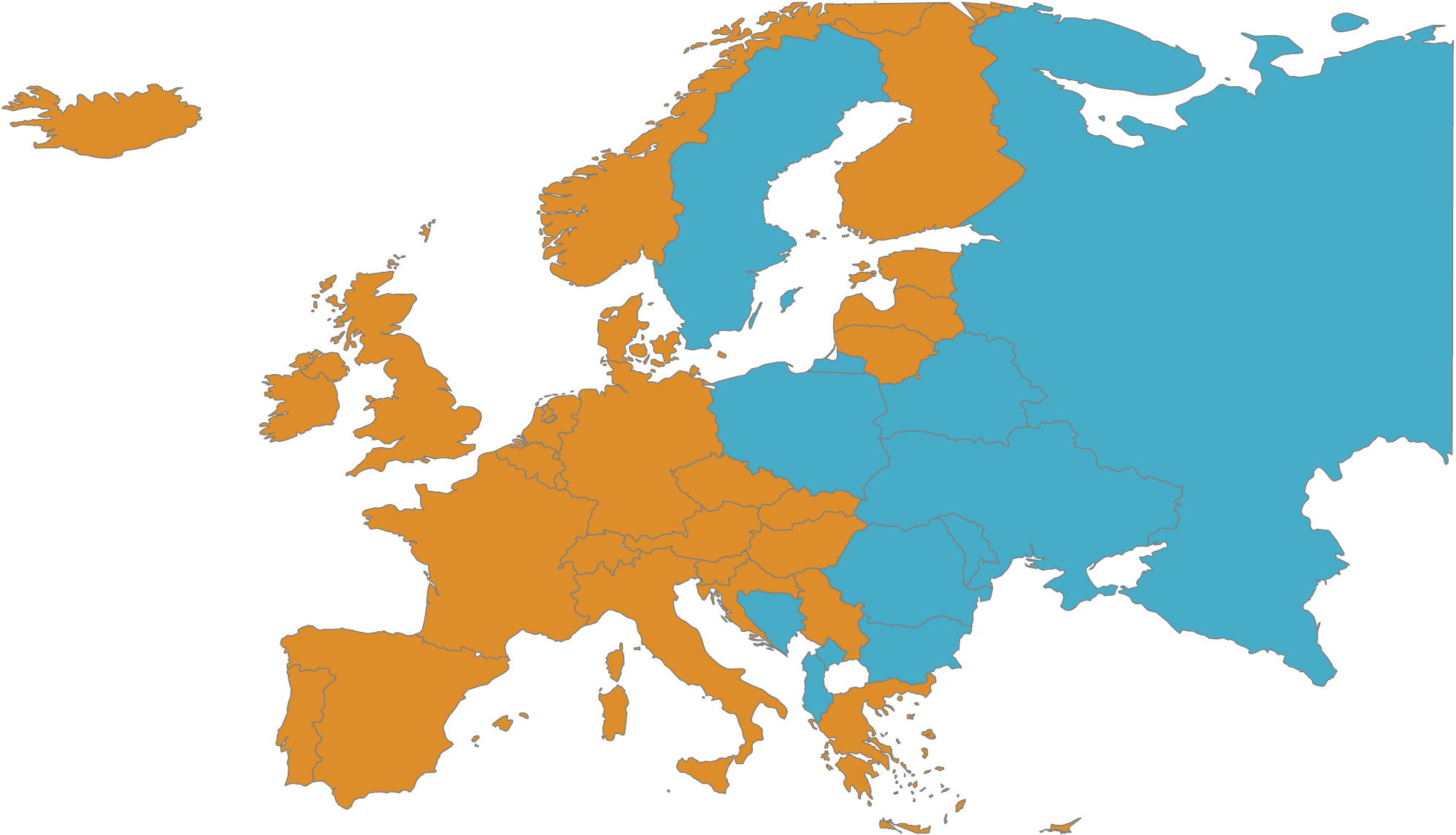}} & & \\
& & \\
& 1 & log \% of cases  per 1000 inhabitants \\
&  2 & log \% of deaths  per 1000 inhabitants \\
&  3 & Stringency index, Government response index  \\
& & \\
& & \\
& & \\
\end{tabular}
\end{center}
\end{table}

%%%%%%%%%%%%%%%%%%%%%%%%%%%%%%%%%%%%%%%%%%%%%%%%%%%%%%

\section{Conclusions}\label{sec:chcharles_conclusions}

%Multivariate time-dependent data can be suitably arranged in three-way structures where each layer introduces its own  peculiar characteristics. When exploring appropriate modelling strategies, it is  required to account for heterogeneous subjects, relations among variables and correlation across different time instants. 

Modeling multivariate time-dependent data requires accounting for heterogeneity among subjects, capturing similarities and differences among variables, as well as correlations between repeated measures. 
Our work has tackled 
%aimed to answer to 
these challenges by proposing a new parametric co-clustering methodology, recasting the widely known Latent Block Model in a time-dependent fashion. The co-clustering model, by simultaneously searching for row and column clusters, partitions three-way matrices in blocks of homogeneous curves. Such approach % seems particularly reasonable in the considered framework since it 
takes into account  the mentioned features of the data while building parsimonious and meaningful summaries.  
As a data generative mechanism for a single curve, we have considered the \emph{Shap Invariant Model} that has turned out to be particularly flexible when embedded in a co-clustering context. The model allows to describe arbitrary time evolution patterns while adequately capturing dependencies among different temporal instants. 
The proposed method compares favorably with the few existing competitors for the simultaneous clustering of subjects and variables with time-dependent data. Our proposal, while producing co-partitions with comparable quality as measured by objective criteria, applies to both functional and longitudinal data, and has relevant advantages in terms of interpretability. The 
option
%chance 
of ``switching off'' some of the random effects, although in principle simplifying the model structure, increases its flexibility, as it allows to encompass different concepts of cluster possibly depending on the specific applications and on subject-matter considerations.  

While further analyses are required to increase our understanding about the general performance of the proposed model, its application to both simulated and real data has provided overall satisfactory results and highlighted some aspects which worth further investigation. 
%Among them, we shall introduce the idea of the \emph{curse of flexibility}, as in some specific situations, the considered specification may induce a degree of flexibility possibly entailing issues from a computational point of view and in the obtained results. 
One interesting direction for future research 
%developments consists in 
is studying possible alternatives to the ICL to be used as selection tools when the model specification in the LBM framework involves random effects. Moreover, alternative choices, for example, for specifying the block mean curves, could be considered and compared with the choices adopted here. A further direction for future work would be
% consist  in 
exploring 
% the chance to resort to 
a fully Bayesian approach to model specification and estimation, possibly handling more easily the random parameters in the model.
%%%%%%%%%%%%%%%%%%%%%%%%%%%%%%%%%%%%%%%%%%%%%%%%%%

%%%%%%%%%%%%%%%%%%%%%%%%%%%%%%%%%%%%%%%%%%%%%%%%%%

\newpage
\bibliographystyle{plainnat}
\bibliography{biblio}

\begin{thebibliography}{52}
\providecommand{\natexlab}[1]{#1}
\providecommand{\url}[1]{\texttt{#1}}
\expandafter\ifx\csname urlstyle\endcsname\relax
  \providecommand{\doi}[1]{doi: #1}\else
  \providecommand{\doi}{doi: \begingroup \urlstyle{rm}\Url}\fi

\bibitem[Anderlucci and Viroli(2015)]{anderlucci2015covariance}
L.~Anderlucci and C.~Viroli.
\newblock Covariance pattern mixture models for the analysis of multivariate
  heterogeneous longitudinal data.
\newblock \emph{The Annals of Applied Statistics}, 9\penalty0 (2):\penalty0
  777--800, 2015.

\bibitem[Ben~Slimen et~al.(2018)Ben~Slimen, Allio, and
  Jacques]{slimen2018model}
Y.S. Ben~Slimen, S.~Allio, and J.~Jacques.
\newblock Model-based co-clustering for functional data.
\newblock \emph{Neurocomputing}, 291:\penalty0 97--108, 2018.

\bibitem[Biernacki et~al.(2000)Biernacki, Celeux, and
  Govaert]{biernacki2000assessing}
C.~Biernacki, G.~Celeux, and G.~Govaert.
\newblock Assessing a mixture model for clustering with the integrated
  completed likelihood.
\newblock \emph{IEEE transactions on pattern analysis and machine
  intelligence}, 22\penalty0 (7):\penalty0 719--725, 2000.

\bibitem[Bouveyron and Jacques(2011)]{bouveyron2011model}
C.~Bouveyron and J.~Jacques.
\newblock Model-based clustering of time series in group-specific functional
  subspaces.
\newblock \emph{Advances in Data Analysis and Classification}, 5\penalty0
  (4):\penalty0 281--300, 2011.

\bibitem[Bouveyron et~al.(2015)Bouveyron, C{\^o}me, and
  Jacques]{bouveyron2015discriminative}
C.~Bouveyron, E.~C{\^o}me, and J.~Jacques.
\newblock The discriminative functional mixture model for a comparative
  analysis of bike sharing systems.
\newblock \emph{The Annals of Applied Statistics}, 9\penalty0 (4):\penalty0
  1726--1760, 2015.

\bibitem[Bouveyron et~al.(2018)Bouveyron, Bozzi, Jacques, and
  Jollois]{bouveyron2018functional}
C.~Bouveyron, L.~Bozzi, J.~Jacques, and F.X. Jollois.
\newblock The functional latent block model for the co-clustering of
  electricity consumption curves.
\newblock \emph{Journal of the Royal Statistical Society: Series C (Applied
  Statistics)}, 67\penalty0 (4):\penalty0 897--915, 2018.

\bibitem[Bouveyron et~al.(2019)Bouveyron, Celeux, Murphy, and
  Raftery]{bouveyron2019model}
C.~Bouveyron, G.~Celeux, T.B. Murphy, and A.E. Raftery.
\newblock \emph{Model-Based Clustering and Classification for Data Science:
  With Applications in R}.
\newblock Cambridge University Press, 2019.

\bibitem[Bouveyron et~al.(2020)Bouveyron, Jacques, Schmutz, Simoes, and
  Bottini]{bouveyron2020co}
C.~Bouveyron, J.~Jacques, A.~Schmutz, F.~Simoes, and S.~Bottini.
\newblock Co-clustering of multivariate functional data for the analysis of air
  pollution in the south of france.
\newblock \emph{HAL preprint hal-02862177}, 2020.

\bibitem[Brauner et~al.(2021)Brauner, Mindermann, Sharma, Johnston, Salvatier,
  Gaven{\v{c}}iak, Stephenson, Leech, Altman, Mikulik,
  et~al.]{brauner2021inferring}
J.M. Brauner, S.~Mindermann, M.~Sharma, D.~Johnston, J.~Salvatier,
  T.~Gaven{\v{c}}iak, A.B. Stephenson, G.~Leech, G.~Altman, V.~Mikulik, et~al.
\newblock Inferring the effectiveness of government interventions against
  {COVID}-19.
\newblock \emph{Science}, 371\penalty0 (6531), 2021.

\bibitem[Corneli and Erosheva(2020)]{corneli2020bayesian}
M.~Corneli and E.~Erosheva.
\newblock A {B}ayesian approach for clustering and exact finite-sample model
  selection in longitudinal data mixtures.
\newblock \emph{HAL preprint hal-02310069v2}, 2020.

\bibitem[Corneli et~al.(2020)Corneli, Bouveyron, and Latouche]{corneli2019co}
M.~Corneli, C.~Bouveyron, and P.~Latouche.
\newblock Co-clustering of ordinal data via latent continuous random variables
  and not missing at random entries.
\newblock \emph{Journal of Computational and Graphical Statistics}, 29\penalty0
  (4):\penalty0 771--785, 2020.

\bibitem[De~Boor(1978)]{de1978practical}
C.~De~Boor.
\newblock \emph{A practical guide to splines}.
\newblock Springer-Verlag, New York, 1978.

\bibitem[De~la Cruz-Mes{\'\i}a et~al.(2008)De~la Cruz-Mes{\'\i}a, Quintana, and
  Marshall]{de2008model}
R.~De~la Cruz-Mes{\'\i}a, F.~A Quintana, and G.~Marshall.
\newblock Model-based clustering for longitudinal data.
\newblock \emph{Computational Statistics \& Data Analysis}, 52\penalty0
  (3):\penalty0 1441--1457, 2008.

\bibitem[Delattre et~al.(2014)Delattre, Lavielle, and
  Poursat]{delattre2014note}
M.~Delattre, M.~Lavielle, and M.~Poursat.
\newblock A note on {BIC} in mixed-effects models.
\newblock \emph{Electronic journal of statistics}, 8\penalty0 (1):\penalty0
  456--475, 2014.

\bibitem[Dempster et~al.(1977)Dempster, Laird, and Rubin]{dempster1977maximum}
A.P. Dempster, Nan~M. Laird, and D.B. Rubin.
\newblock Maximum likelihood from incomplete data via the em algorithm.
\newblock \emph{Journal of the Royal Statistical Society: Series B
  (Methodological)}, 39\penalty0 (1):\penalty0 1--22, 1977.

\bibitem[Diggle et~al.(2002)Diggle, Heagerty, Liang, Heagerty, and
  Zeger]{diggle2002analysis}
P.J. Diggle, P.~Heagerty, K.Y. Liang, P.J. Heagerty, and S.~Zeger.
\newblock \emph{Analysis of longitudinal data}.
\newblock Oxford University Press, 2002.

\bibitem[Erosheva et~al.(2014)Erosheva, Matsueda, and
  Telesca]{erosheva2014breaking}
E.~Erosheva, R.L. Matsueda, and D.~Telesca.
\newblock Breaking bad: Two decades of life-course data analysis in
  criminology, developmental psychology, and beyond.
\newblock \emph{Annual Review of Statistics and Its Application}, 1:\penalty0
  301--332, 2014.

\bibitem[Flaxman et~al.(2020)Flaxman, Mishra, Gandy, Unwin, Mellan, Coupland,
  Whittaker, Zhu, Berah, Eaton, et~al.]{flaxman2020estimating}
S.~Flaxman, S.~Mishra, A.~Gandy, H.J.T. Unwin, T.A. Mellan, H.~Coupland,
  C.~Whittaker, H.~Zhu, T.~Berah, J.W. Eaton, et~al.
\newblock Estimating the effects of non-pharmaceutical interventions on
  {COVID}-19 in europe.
\newblock \emph{Nature}, 584\penalty0 (7820):\penalty0 257--261, 2020.

\bibitem[Fraley and Raftery(2002)]{fraley2002model}
C.~Fraley and A.E. Raftery.
\newblock Model-based clustering, discriminant analysis, and density
  estimation.
\newblock \emph{Journal of the American statistical Association}, 97\penalty0
  (458):\penalty0 611--631, 2002.

\bibitem[Fr{\"u}hwirth-Schnatter(2011)]{frauhwirth2011model}
S.~Fr{\"u}hwirth-Schnatter.
\newblock Panel data analysis: a survey on model-based clustering of time
  series.
\newblock \emph{Advances in Data Analysis and Classification}, 5\penalty0
  (4):\penalty0 251--280, 2011.

\bibitem[Govaert and Nadif(2003)]{govaert2003clustering}
G.~Govaert and M.~Nadif.
\newblock Clustering with block mixture models.
\newblock \emph{Pattern Recognition}, 36\penalty0 (2):\penalty0 463--473, 2003.

\bibitem[Govaert and Nadif(2008)]{govaert2008block}
G.~Govaert and M.~Nadif.
\newblock Block clustering with bernoulli mixture models: Comparison of
  different approaches.
\newblock \emph{Computational Statistics \& Data Analysis}, 52\penalty0
  (6):\penalty0 3233--3245, 2008.

\bibitem[Govaert and Nadif(2010)]{govaert2010latent}
G.~Govaert and M.~Nadif.
\newblock Latent block model for contingency table.
\newblock \emph{Communications in Statistics - Theory and Methods}, 39\penalty0
  (3):\penalty0 416--425, 2010.

\bibitem[Govaert and Nadif(2013)]{govaert2013co}
G.~Govaert and M.~Nadif.
\newblock \emph{Co-clustering: models, algorithms and applications}.
\newblock John Wiley \& Sons, 2013.

\bibitem[Hale et~al.(2020)Hale, Angrist, Cameron-Blake, Hallas, Kira, Majumdar,
  Petherick, Tatlow, and Webster]{covid}
T.~Hale, N.~Angrist, E.~Cameron-Blake, L.~Hallas, B.~Kira, S.~Majumdar,
  T.~Petherick, A.~Phillips, H.~Tatlow, and S.~Webster.
\newblock \emph{Oxford COVID-19 Government Response Tracker, Blavatnik School
  of Government}, 2020.
\newblock URL
  \url{https://www.bsg.ox.ac.uk/research/research-projects/coronavirus-government-
  response-tracker}.

\bibitem[Harring and Liu(2016)]{harring2016comparison}
J.R. Harring and J.~Liu.
\newblock A comparison of estimation methods for nonlinear mixed-effects models
  under model misspecification and data sparseness: A simulation study.
\newblock \emph{Journal of Modern Applied Statistical Methods}, 15\penalty0
  (1):\penalty0 27, 2016.

\bibitem[Hubert and Arabie(1985)]{hubert1985comparing}
L.~Hubert and P.~Arabie.
\newblock Comparing partitions.
\newblock \emph{Journal of Classification}, 2\penalty0 (1):\penalty0 193--218,
  1985.

\bibitem[Jacques and Biernacki(2018)]{jacques2018model}
J.~Jacques and C.~Biernacki.
\newblock Model-based co-clustering for ordinal data.
\newblock \emph{Computational Statistics \& Data Analysis}, 123:\penalty0
  101--115, 2018.

\bibitem[Jacques and Preda(2014)]{jacques2014functional}
J.~Jacques and C.~Preda.
\newblock Functional data clustering: a survey.
\newblock \emph{Advances in Data Analysis and Classification}, 8\penalty0
  (3):\penalty0 231--255, 2014.

\bibitem[Keribin et~al.(2015)Keribin, Brault, Celeux, and
  Govaert]{keribin2015estimation}
C.~Keribin, V.~Brault, G.~Celeux, and G.~Govaert.
\newblock Estimation and selection for the latent block model on categorical
  data.
\newblock \emph{Statistics and Computing}, 25\penalty0 (6):\penalty0
  1201--1216, 2015.

\bibitem[Lawton et~al.(1972)Lawton, Sylvestre, and Maggio]{lawton1972self}
W.H. Lawton, E.A. Sylvestre, and M.S. Maggio.
\newblock Self modeling nonlinear regression.
\newblock \emph{Technometrics}, 14\penalty0 (3):\penalty0 513--532, 1972.

\bibitem[Liao(2005)]{liao2005clustering}
T.W. Liao.
\newblock Clustering of time series data - a survey.
\newblock \emph{Pattern recognition}, 38\penalty0 (11):\penalty0 1857--1874,
  2005.

\bibitem[Lindstrom(1995)]{lindstrom1995self}
M.J. Lindstrom.
\newblock Self-modelling with random shift and scale parameters and a free-knot
  spline shape function.
\newblock \emph{Statistics in Medicine}, 14\penalty0 (18):\penalty0 2009--2021,
  1995.

\bibitem[Lindstrom and Bates(1990)]{lindstrom1990nonlinear}
M.J. Lindstrom and D.~Bates.
\newblock Nonlinear mixed effects models for repeated measures data.
\newblock \emph{Biometrics}, 46\penalty0 (3):\penalty0 673--687, 1990.

\bibitem[Lomet(2012)]{lomet2012selection}
A.~Lomet.
\newblock \emph{S{\'e}lection de mod{\`e}le pour la classification crois{\'e}e
  de donn{\'e}es continues}.
\newblock PhD thesis, Compi{\`e}gne, 2012.

\bibitem[McNicholas and Murphy(2010)]{mcnicholas2010model}
P.D. McNicholas and T.B. Murphy.
\newblock Model-based clustering of longitudinal data.
\newblock \emph{Canadian Journal of Statistics}, 38\penalty0 (1):\penalty0
  153--168, 2010.

\bibitem[Nagin(2009)]{nagin2009group}
D.~Nagin.
\newblock \emph{Group-based modeling of development}.
\newblock Harvard University Press, 2009.

\bibitem[Pinheiro and Bates(2006)]{pinheiro2006mixed}
J.~Pinheiro and D.~Bates.
\newblock \emph{Mixed-effects models in S and S-PLUS}.
\newblock Springer Science \& Business Media, 2006.

\bibitem[Pinheiro et~al.(2019)Pinheiro, Bates, DebRoy, Sarkar, and {R Core
  Team}]{nlmepack}
J.~Pinheiro, D.~Bates, S.~DebRoy, D.~Sarkar, and {R Core Team}.
\newblock \emph{{nlme}: Linear and Nonlinear Mixed Effects Models}, 2019.
\newblock URL \url{https://CRAN.R-project.org/package=nlme}.
\newblock R package version 3.1-139.

\bibitem[{R Core Team}(2019)]{rcore}
{R Core Team}.
\newblock \emph{R: A Language and Environment for Statistical Computing}.
\newblock R Foundation for Statistical Computing, Vienna, Austria, 2019.
\newblock URL \url{https://www.R-project.org/}.

\bibitem[Ramsay and Li(1998)]{ramsay1998curve}
J.O. Ramsay and X.~Li.
\newblock Curve registration.
\newblock \emph{Journal of the Royal Statistical Society: Series B
  (Methodological)}, 60\penalty0 (2):\penalty0 351--363, 1998.

\bibitem[Ramsay and Silverman(2005)]{ramsey2005functional}
J.O. Ramsay and B.W. Silverman.
\newblock \emph{Functional data analysis}.
\newblock Springer, New York, 2005.

\bibitem[Rice(2004)]{rice2004functional}
J.A. Rice.
\newblock Functional and longitudinal data analysis: perspectives on smoothing.
\newblock \emph{Statistica Sinica}, pages 631--647, 2004.

\bibitem[Robert et~al.(2020)Robert, Vasseur, and Brault]{robert2020comparing}
V.~Robert, Y.~Vasseur, and V.~Brault.
\newblock Comparing high-dimensional partitions with the co-clustering adjusted
  rand index.
\newblock \emph{Journal of Classification}, pages 1--29, 2020.

\bibitem[Selosse et~al.(2020)Selosse, Jacques, and Biernacki]{selosse2020model}
M.~Selosse, J.~Jacques, and C.~Biernacki.
\newblock Model-based co-clustering for mixed type data.
\newblock \emph{Computational Statistics \& Data Analysis}, 144:\penalty0
  106866, 2020.

\bibitem[Telesca and Inoue(2008)]{telesca2008bayesian}
D.~Telesca and L.Y.T. Inoue.
\newblock Bayesian hierarchical curve registration.
\newblock \emph{Journal of the American Statistical Association}, 103\penalty0
  (481):\penalty0 328--339, 2008.

\bibitem[Telesca et~al.(2012)Telesca, Erosheva, Kreager, and
  Matsueda]{telesca2012modeling}
D.~Telesca, E.~Erosheva, D.~A Kreager, and R.L. Matsueda.
\newblock Modeling criminal careers as departures from a unimodal population
  age--crime curve: the case of marijuana use.
\newblock \emph{Journal of the American Statistical Association}, 107\penalty0
  (500):\penalty0 1427--1440, 2012.

\bibitem[Viroli(2011{\natexlab{a}})]{viroli2011finite}
C.~Viroli.
\newblock Finite mixtures of matrix normal distributions for classifying
  three-way data.
\newblock \emph{Statistics and Computing}, 21\penalty0 (4):\penalty0 511--522,
  2011{\natexlab{a}}.

\bibitem[Viroli(2011{\natexlab{b}})]{viroli2011model}
C.~Viroli.
\newblock Model based clustering for three-way data structures.
\newblock \emph{Bayesian Analysis}, 6\penalty0 (4):\penalty0 573--602,
  2011{\natexlab{b}}.

\bibitem[Wu(1983)]{wu1983convergence}
C.J. Wu.
\newblock On the convergence properties of the em algorithm.
\newblock \emph{The Annals of statistics}, pages 95--103, 1983.

\bibitem[Wyse and Friel(2012)]{wyse2012block}
J.~Wyse and N.~Friel.
\newblock Block clustering with collapsed latent block models.
\newblock \emph{Statistics and Computing}, 22\penalty0 (2):\penalty0 415--428,
  2012.

\bibitem[Wyse et~al.(2017)Wyse, Friel, and Latouche]{wyse2017inferring}
J.~Wyse, N.~Friel, and P.~Latouche.
\newblock Inferring structure in bipartite networks using the latent blockmodel
  and exact {ICL}.
\newblock \emph{Network Science}, 5\penalty0 (1):\penalty0 45--69, 2017.

\end{thebibliography}

\end{document}